\documentclass[a4paper,11pt]{article}
\pdfoutput=1 %

\usepackage{jcappub} %

\usepackage[T1]{fontenc} %
\usepackage{physics}
\usepackage[separate-uncertainty=true]{siunitx}
\usepackage[normalem]{ulem}
\usepackage{nicefrac}
\usepackage{comment}
\usepackage{slashed}
\usepackage{xcolor}
\usepackage{cancel}
\begin{document}
\title{\boldmath Forbidden dark matter annihilation into leptons with full collision terms}

\author[a]{Amin Aboubrahim,}
\author[a]{Michael Klasen,}
\author[a]{and Luca Paolo Wiggering}

\affiliation[a]{Institut für Theoretische Physik, Westfälische Wilhelms-Universität Münster,\\Wilhelm-Klemm-Straße 9, 48149 Münster,  Germany}

\emailAdd{aabouibr@uni-muenster.de}
\emailAdd{michael.klasen@uni-muenster.de}
\emailAdd{luca.wiggering@uni-muenster.de}

\abstract{
The standard approach of calculating the relic density of thermally produced dark matter based on the assumption of kinetic equilibrium is known to fail for forbidden dark matter models since only the high momentum tail of the dark matter phase space distribution function contributes significantly to dark matter annihilations. Furthermore, it is known that the computationally less expensive Fokker-Planck approximation for the collision term describing elastic scattering processes between non-relativistic dark matter particles and the Standard Model thermal bath breaks down if both scattering partners are close in mass. This, however, is the defining feature of the forbidden dark matter paradigm. In this paper, we therefore include the full elastic collision term in the full momentum-dependent Boltzmann equation as well as in a set of fluid equations that couple the evolution of the number density and dark matter temperature for a simplified model featuring forbidden dark matter annihilations into muon or tau leptons through a scalar mediator. On the technical side, we perform all angular integrals in the full collision term analytically and take into account the effect of dark matter self-interactions on the relic density. The overall phenomenological outcome is that the updated relic density calculation results in a significant reduction of the experimentally allowed parameter space compared to the traditional approach, which solves only for the abundance. In addition, almost the entire currently viable parameter space can be probed with CMB-S4, next-generation beam-dump experiments or at a future high-luminosity electron-position collider, except for the resonant region where the mediator corresponds to approximately twice the muon or tau mass. 
}

\begin{flushright}
{\large \tt MS-TP-23-15}
\end{flushright}

\maketitle

\section{Introduction}
The existence of dark matter (DM) is confirmed by numerous astrophysical observations~\cite{Klasen:2015uma,Freese:2017idy} and the associated energy density has been determined precisely to the value $\Omega_\chi h^2 = \SI{0.12\pm0.001}{}$
through a series of analyses of the Cosmic Microwave Background (CMB) within the $\Lambda$CDM model \cite{Planck:2018vyg}. Nevertheless, the nature and intrinsic properties of DM remain unknown. A theoretically appealing explanation is that all of the observed DM consists of a single new elementary particle species with sufficiently strong interactions with the Standard Model (SM) to have established full thermal equilibrium at some point in the early Universe. In this picture, today's DM relic abundance is set once the DM annihilation cross section falls below the Hubble expansion rate so that the dark sector drops out of chemical equilibrium and the DM number density becomes an effective comoving constant (\emph{freezes out}). Within the usual approach to determine the relic density, the classical Boltzmann equation describing the evolution of the DM phase space distribution function in a Friedmann-Robertson-Walker (FRW) Universe is solved. To simplify the calculation, the majority of numerical DM codes, e.g.~\cite{Belanger:2006is,Arbey:2009gu,Bringmann:2018lay,Ambrogi:2018jqj}, assume that kinetic equilibrium holds until long after chemical decoupling which then allows to trace only the integral over the DM phase space distribution function (the number density) \cite{Gondolo:1990dk,Edsjo:1997bg}.  However, kinetic decoupling might occur much earlier than chemical decoupling even in simple models given that the DM annihilation cross section exhibits a strong velocity dependence caused by e.g. resonances, thresholds or the Sommerfeld enhancement effect \cite{Binder:2017rgn,Abe:2020obo,Abe:2021jcz,Binder:2021bmg}. As a consequence, the final value of the relic abundance can be altered by more than an order of magnitude compared to the traditional number density approach~\cite{Binder:2017rgn,Binder:2021bmg}. In order to adequately model the effect of early kinetic decoupling,
extensions of this ``standard'' number density Boltzmann equation (nBE) approach have been developed. Possibilities are for example $(1)$ to solve a set of coupled Boltzmann equations (cBE) assuming that deviations from equilibrium are entirely described by the chemical potential and the temperature or $(2)$ to obtain a numerical solution for the full Boltzmann equation (fBE) at the level of the phase space distribution function. Compared to the nBE treatment, the fluid approximation consists of two coupled Boltzmann equations, one for the number density and one for the velocity dispersion (``the DM temperature'') by keeping the assumption of a thermal DM distribution, but at a temperature different from the photon temperature. It should be noted that the same kind of hydrodynamical formalism is also used to estimate the mass of the smallest dark matter subhalos~\cite{vandenAarssen:2012ag} and to model the dynamics of domain walls within cosmological phase transitions \cite{Moore:1995si}. Both of these two methods are available in the publicly available and \texttt{Wolfram Language} based numerical precision code \texttt{DRAKE}~\cite{Binder:2021bmg}. However, the default implementation of the elastic collision term for both approaches relies on a Fokker-Planck (FP) type operator derived under the assumption of a small-momentum transfer between DM and SM particles in the thermal bath compared to the average DM momentum which is not necessarily the case if the DM particle and the scattering partner are close in mass. This, however, is a defining feature of forbidden or sub-threshold DM, which is a class of models where DM dominantly annihilates into heavier states~\cite{Griest:1990kh,DAgnolo:2015ujb,Delgado:2016umt}. These annihilations are made possible through the sufficiently large temperatures in the early Universe. 

Going beyond the current state of research described above, we design our own $\texttt{C}$-based Boltzmann equation solver with full elastic collision terms for both approaches and analyze, as an example, the forbidden DM model~\cite{DAgnolo:2020mpt} in which a singlet Dirac DM particle couples to SM leptons via a new scalar mediator. We stress again that the analysis is carried out at the level of the phase space density without relying on simplifying approximations of the elastic collision term or on the Fokker-Planck version of the cBE approach alone as in Ref.~\cite{Liu:2023kat}. For this purpose, we perform the angular integrals of the full elastic collision term analytically. This calculation and the associated methodology respond to the increasing interest in full solutions of the momentum-dependent Boltzmann equation not only in the context of the DM relic density~\cite{Binder:2016pnr,Ala-Mattinen:2019mpa,Brummer:2019inq,Binder:2021bmg,Du:2021jcj,Ala-Mattinen:2022nuj,Hryczuk:2022gay}, but also in many other areas, e.g. the precise computation of the effective number of neutrino species in the early Universe \cite{Hannestad:2015tea}, leptogenesis \cite{Hahn-Woernle:2009jyb}, cosmic inflation \cite{Ghosh:2022hen} and gravitational waves from first order phase transitions \cite{DeCurtis:2022hlx}. As an alternative to the Boltzmann framework, Langevin simulations have been proposed to deal with non-equilibrium momentum distributions of non-relativistic DM in a FRW background~\cite{Kim:2023bxy}.
 
This paper is organized as follows: in Sec.~\ref{sec:boltzmann} we introduce the momentum-dependent Boltzmann equation as well as the fluid equations and describe the numerical solution strategy. The particle content of the forbidden DM model is introduced in Sec. \ref{sec:forbiddenDM} along with a detailed comparison of the relic density obtained with the different approaches including a discussion of the evolution of the phase space distribution function and the effect from DM self-scattering processes, followed by the presentation of the impact of current and projected limits besides the relic density from CMB observations, beam-dump and collider experiments on the parameter space. Conclusions are given in Sec. \ref{sec:conclusion}. Self-interaction cross sections as well as form factors for loop-induced processes are provided in App.~\ref{app:XSec} and the details of the derivation of the elastic scattering collision term are given in App.~\ref{app:D}. The latter constitutes one of our main results in this work.

\section{From the Boltzmann equation to the relic abundance of thermal DM}\label{sec:boltzmann}

The standard starting point for the calculation of the DM relic density is the classical Boltzmann equation
\begin{equation}
    \pdv{f_\chi}{t} - H p \pdv{f_\chi}{p} = C_{\mathrm{coll}}[f_\chi],
    \label{eq:Boltz1}
\end{equation}
describing the evolution of the phase space distribution function $f_\chi(p,t)$ of a DM particle $\chi$ with momentum $p$ and corresponding energy $E=\sqrt{p^2 + m_\chi^2}$ in time $t$ within the FRW cosmological model, where $H=\dot{a}/a$ denotes the Hubble expansion rate and $a(t)$ the scale factor~\cite{bernstein_1988}. The collision term $C_{\mathrm{coll}}[f_\chi]$ takes into account the loss and gain of DM particles at the momentum $p$ through interactions with the SM and is defined in detail in the next section. The momentum derivative responsible for modeling the cosmological redshift can be absorbed into the phase space density by rewriting the Boltzmann equation in terms of a dimensionless comoving momentum $q \sim p a$, thus turning the partial differential equation into an infinite set of coupled ordinary differential equations
\begin{equation}    
    \dv{f_\chi(q,t)}{t} =  C_{\mathrm{coll}}[f_\chi],
\end{equation}
where one has to make the identification $f_\chi(p,t) = f_\chi(q,t)$. Since the thermal equilibrium distribution functions $f_{\pm}(E) = 1/(1 \pm e^{-E/T})$ for a Fermi-Dirac/Bose-Einstein distribution or $f_{\mathrm{MB}}(E) = e^{-E/T}$ for a Maxwell-Boltzmann distribution 
depend only implicitly on time through the temperature, it is instructive to express the evolution of $f_\chi$  not terms of time but through the photon temperature $T$ or alternatively through the dimensionless quantity $x=m_0/T$, where $m_0$ is in general some reference scale which we set to the DM mass $m_\chi$.
This is achieved by using the scale factor as a time variable in the intermediate step $\dv{f_\chi}{a} = \frac{1}{H a} C_{\mathrm{coll}}[f_\chi]$ and replacing the derivative with respect to $a(t)$ afterwards by assuming entropy conservation $\dv{x} (s a^3)=0$ giving
\begin{equation}
     \dv{f_\chi}{x} = -\dv{s}{x} \frac{1}{3 s H} C_{\mathrm{coll}}[f_\chi].
\end{equation}
During radiation domination, i.e. for $T \gtrsim \SI{100}{\electronvolt}$, the entropy density $s = h_\mathrm{eff}(T) \frac{2 \pi^2}{45} T^3$ and the energy density $\rho = g_\mathrm{eff}(T) \frac{\pi^2}{30} T^4$ can be safely expressed in terms of the SM effective number of degrees of freedom, $h_\mathrm{eff}(T)$ and $ g_\mathrm{eff}(T)$, for which we use the tabulated values from the lattice QCD calculation by \emph{Drees et al.} \cite{Drees:2015exa}. 
Applying the Friedmann equation for a flat Universe $H^2 = \rho/(3 M^2_\mathrm{Pl})$ with the reduced Planck mass $M_\mathrm{Pl}=1/\sqrt{8 \pi G}$
gives the final equation
\begin{equation}
     \dv{f_\chi(q,x)}{x} = \sqrt{\frac{90}{\pi^2}} \frac{\tilde{g}(T)}{ g^{1/2}_\mathrm{eff}(T)} \frac{x M_{\mathrm{Pl}}}{m_0^2}  C_{\mathrm{coll}}[f_\chi],\label{eq:boltzFinal}
\end{equation}
where we have introduced the shorthand notation
\begin{equation}
    \tilde{g}(T) = 1 + \frac{1}{3} \dv{\ln h_\mathrm{eff}(T)}{\ln T}.
\end{equation}
Note that there already occur differences in the DM abundance at the level of Eq.~\eqref{eq:boltzFinal} due to e.g. uncertainties in the SM equation of state~\cite{Saikawa:2020swg} or unknown additional light degrees of freedom before Big Bang nucleosynthesis which change the expansion history of the Universe~\cite{Arbey:2008kv}.
Lastly, it is necessary to pick a convention for the proportionality factor relating the comoving momentum $q$ to the physical momentum $p$. Since it is only possible to consider the ratio of the scale factor relative to its value at some other reference temperature $T'$, we define the comoving momentum to be
\begin{equation}
    q =\left[h_{\mathrm{eff}}(T')\right]^{-1/3} \frac{p }{T'}  \frac{a(T)}{a(T')} = \left[h_{\mathrm{eff}}(T)\right]^{-1/3} \frac{p}{T}.
\end{equation}

\subsection{The collision terms}
As a general expression for the collision operator is rather cumbersome, we provide as the only example relevant to this work the collision term for a generic two-particle interaction $a b \leftrightarrow 12$ in a $CP$-conserving theory
\begin{multline}
    C_{\mathrm{coll}}[f_a] = \frac{1}{2 E_a g_a} \int \dd{\Pi_b}\dd{\Pi_1}\dd{\Pi_2}  |\mathcal{M}_{a b\to 1 2}|^2 \\
   \times (2\pi)^4 \delta^{(4)}(p_a + p_b - p_1 - p_2) \,\mathcal{P}(f_a,f_b,f_1,f_2),
\end{multline}
where $a$ carries $g_a$ internal degrees of freedom and $\dd{\Pi_i} = \dd[3]{p_i}/[(2\pi)^32 E_i]$ is the Lorentz invariant integration measure. In our convention $|\mathcal{M}_{a b\to 1 2}|^2=|\mathcal{M}_{1 2\to a b}|^2$ always represents the squared matrix element \emph{summed} (not averaged) over both initial and final internal degrees of freedom and includes a symmetry factor $\nicefrac{1}{2}$ for identical particles in either the initial or final state. 
The phase space densities are contained in the population factor 
\begin{equation}
    \mathcal{P}(f_a,f_b,f_1,f_2) = f_1 f_2 (1 \pm f_a)(1 \pm f_b) \\- f_a f_b (1 \pm f_1)(1 \pm f_2),
\end{equation}
with $+(-)$ accounting for Bose enhancement (Pauli blocking) of the final states. The first term on the right-hand-side is often called the gain term, whereas the second is referred to as the loss term. The associated statistical factors can be dropped for a non-relativistic gas, i.e. if $1 \pm f \approx 1$. According to the number of appearances of the unknown distribution function $f_\chi$ in the loss and gain terms, the collision term can be split further into a number changing contribution, $C_{\mathrm{ann}}$, from DM annihilation processes into SM final states, a number conserving part, $C_{\mathrm{el}}$, from elastic scattering processes of DM with the SM thermal bath responsible for maintaining kinetic equilibrium and a collision term $C_{\mathrm{self}}$ describing self-scattering processes. In the following part, the relevant individual contributions to $C_{\mathrm{coll}}$ are worked out in more detail. 

For non-relativistic DM, the Bose enhancement and Pauli blocking factors can be safely neglected, thus implying that in $C_{\mathrm{ann}}$, SM particles are, due to energy conservation, in very good approximation described by a Boltzmann distribution and that consequently the statistical factors accompanying the SM densities can be dropped as well. These simplifications allow to express the annihilation collision term as
\begin{equation}
   C_{\mathrm{ann}}[f_\chi] = \frac{g_\chi}{2\pi^2 } \int_0^\infty \dd{p_b} p_b^2 \langle  v_{\text{M\o l}} \sigma_{\mathrm{ann}}  \rangle_{\theta} \left[f_\mathrm{MB}(p_a) f_\mathrm{MB}(p_b) - f_\chi(p_a) f_\chi(p_b) \right],
   \label{eq:Cann}
\end{equation}
with the azimuthally averaged annihilation cross section 
\begin{equation}
    \langle  v_{\text{M\o l}} \sigma_{\mathrm{ann}} \rangle_{\theta} = \frac{1}{2} \int_{-1}^1 \dd{\!\cos\theta}  v_{\text{M\o l}} \sigma_{\mathrm{ann}}\,. 
\end{equation}
Here the M\o ller velocity is $v_{\text{M\o l}} = \sqrt{s(s - 4 m^2_\chi)}/(2 E_a E_b)$, with $\theta$ denoting the angle between the incoming momenta $p_a$ and $p_b$.

In contrast to the annihilation term, the elastic collision term 
\begin{multline}
    C_{\mathrm{el}}[f_\chi] = \frac{1}{2 E_a g_a} \int \dd{\Pi_b}\dd{\Pi_1}\dd{\Pi_2} |\mathcal{M}_{\chi j \to \chi j}|^2  (2\pi)^4 \delta^{(4)}(p_a + p_b - p_1 - p_2)  \\ \times \left[f_\chi(p_1) f^{(j)}_\pm(p_2) (1 \mp f^{(j)}_\pm(p_b))  - f_\chi(p_a) f^{(j)}_{\pm}(p_b)(1 \mp f^{(j)}_\pm(p_2)\right],
\end{multline}
contains the unknown DM distribution function $f_\chi$ along with the equilibrium density $f^{(j)}_\pm$ of a SM particle $j$ in both the loss and gain terms, significantly increasing the evaluation complexity of $C_{\mathrm{el}}$ 
since there remain in general four integrals after imposing four-momentum conservation and using the rotational symmetry around the axis defined by the incoming momentum. Such a general, yet numerically very expensive, parametrization of the collision term for generic two-particle interactions has been put forward by \emph{Hannestad et al.} \cite{Hannestad:1995rs} and has been, to the best of our knowledge, the method of choice so far to evaluate the full collision term in the context of predicting the DM relic abundance, see e.g. \cite{Du:2021jcj,Ala-Mattinen:2022nuj,Hryczuk:2022gay}. 

In contrast to the method usually used in the literature,  
we provide, as an integral part of this work, a parametrization of the general $2$-to-$2$ collision term along with an integration scheme which together allow to 
perform seven out of the nine momentum integrals analytically for the particular case where the matrix element depends only on one Mandelstam variable whose spatial component we label as $k^2$ and that can be brought into the form 
\begin{equation}
    |\mathcal{M}_{ab\to 1 2}|^2 = c_0 + \frac{c_1}{\Delta_1 - k^2} + \frac{c_2}{(\Delta_2 - k^2)^2},
\end{equation}
with $k$-independent coefficients $c_0$, $c_1$, $c_2$ and two other free parameters $\Delta_1$, $\Delta_2$. In the end, the collision operator contains the integration kernel $\Pi$ and takes the form
\begin{align}
     C_{\mathrm{coll}}[f_a] = \frac{1}{128 \pi^3 E_a  p_a } \int^\infty_{m_1} \dd{E_1} \int^\infty_{\max(m_b,E_1-E_a+m_2)} \dd{E_b}  \,\Pi(E_a,E_b,E_1)\,\mathcal{P}(f_a,f_b,f_1,f_2),
     \label{eq:Ccoll}
\end{align}
thus requiring the same, or even less, computational effort as the annihilation term in Eq.~\eqref{eq:Cann}.
More details on how Eq.~\eqref{eq:Ccoll} was derived and an explicit example for the integration kernel are provided in App.~\ref{app:D}. Let us stress that the application of this technique is not limited to elastic scatterings but can also be applied to e.g. the annihilation operator if quantum statistical effects need to be included.

In contrast to the full expression for $C_{\mathrm{el}}$, the DM code \texttt{DRAKE} implements by default the scattering term through the Fokker-Planck type operator \cite{Binder:2016pnr}
\begin{equation}
    C_{\mathrm{FP}}[f_\chi] = \frac{\gamma(T)}{2} \left[T E \partial_p^2 + \left(p + 2 T \frac{E}{p} + T \frac{p}{E}\right) \partial_p + 3\right] f_\chi(p), 
    \label{eq:CFP}
\end{equation}
which is valid under the assumption of non-relativistic DM and if the momentum transfer is small compared to the DM mass.
Here, the particle physics model enters through the momentum exchange rate which is an integral over the energy $\omega = \sqrt{k^2 + m_j^2}$ of the bath particle $j$ and reads
\begin{equation}
    \gamma(T) = \frac{1}{48 \pi^3 g_\chi m_\chi^3 T}  \int_{m_j}^\infty \dd{\omega}  f_{\pm}(\omega)  [1 \mp f_{\pm}(\omega)] \left(k^4 \langle |\mathcal{M}_{\chi j \to \chi j}|^2 \rangle_t \right),
\end{equation}
while the averaged matrix element is given by
\begin{equation}
     \langle |\mathcal{M}_{\chi j \to \chi j}|^2 \rangle_t = \frac{1}{8 k^4} \int^0_{-4 k^2_{\mathrm{cm}}} \dd{t} (-t) |\mathcal{M}_{\chi j \to \chi j}|^2,
\end{equation}
where the lower integration limit is defined through the center-of-mass momentum $k^2_{\mathrm{cm}} = \left( s - (m_\chi - m_j)^2\right)(s - (m_\chi + m_j)^2)/4 s$ evaluated at $s = m_\chi^2 + 2\omega m_\chi + m_j^2$.

\subsection{The fluid dynamics approach}
An alternative to the computationally expensive momentum-dependent Boltzmann equation is to take a hydrodynamic approach and consider averaged quantities 
instead. For DM, the most interesting ones are the DM number density  
\begin{equation}
n_\chi=g_\chi \int \frac{\dd[3]{p}}{(2\pi)^3} f_\chi(p),   
\end{equation}
and the velocity dispersion or DM ``temperature'' $T_\chi$ which one can define through the second moment of $f_\chi$ as
\begin{equation}
     T_\chi = \Bigl\langle \frac{p^2}{3 E} \Bigr\rangle = \frac{g_\chi}{n_\chi} \int \frac{\dd[3]{p}}{(2\pi^3)} \frac{p^2}{3 E} f_\chi(p).
    \label{eq:TChiPS}
\end{equation}
This temperature definition has the advantage of becoming an identity if $f_\chi$ has a Maxwellian shape $f_\chi \sim e^{-E/T_\chi}$.
However, in order to close the Boltzmann hierarchy, this approach requires further assumptions on the phase space distribution function. The simplest and traditional approach is to assume that DM is non-relativistic and therefore obeys the shape of a Maxwell-Boltzmann distribution $f_\chi \sim e^{-E/T}$ with the same temperature as the visible sector. Then, the Boltzmann equation for the number density is obtained by integrating Eq.~\eqref{eq:boltzFinal} over $g_\chi \dd[3]{p}/(2\pi)^{3}$ so that only the contribution from the number changing annihilation collision term survives and one arrives at the often quoted number density Boltzmann equation 
\begin{equation}
    \dv{n_\chi}{t} + 3 H n_\chi = \langle \sigma v\rangle_T \left((n_\chi^{\mathrm{eq}})^2 - n_\chi^2\right),
    \label{eq:nBE}
\end{equation}
with $n_\chi^{\mathrm{eq}} = g_\chi m_\chi^3 K_2(x)/(2 \pi^2 x)$ being the equilibrium number density for a vanishing chemical potential $\mu = 0$. The thermally averaged DM annihilation cross section into SM particles can be stated in terms of a single integral over the collision energy  \cite{Gondolo:1990dk}
\begin{equation}
     \langle \sigma  v \rangle_T =  \frac{1}{8 m_\chi^4 T K_2^2(m_\chi/T)}\int_{4 m_\chi^2}^\infty \dd{s}   \sigma_{\chi\chi \to \mathrm{SM}}(s)\sqrt{s}(s - 4 m_\chi^2) K_1(\sqrt{s}/T),
     \label{eq:vsGG}
\end{equation}
where $K_1$ and $K_2$ are the modified Bessel functions of the second kind and degrees one and
two, respectively. If elastic scattering processes between the dark sector and the SM are not efficient enough to keep both sectors in kinetic equilibrium, one can open up the second moment of the Boltzmann equation and treat it as an equation for the DM temperature $T_\chi$. In the resulting system of equations 
\begin{align}
    &\frac{1}{Y_\chi}\dv{Y_\chi}{x} = \frac{s Y_\chi}{x \tilde{H}} \left[\frac{(Y_\chi^{\mathrm{eq}})^2}{Y_\chi^2} \langle \sigma v \rangle_T  - \langle \sigma v \rangle_{T_\chi}\right],  \\
    & \frac{1}{y}\dv{y}{x} = \frac{\langle C_{\mathrm{el}} \rangle_2}{x \tilde{H}}  + \frac{s Y_\chi}{x \tilde{H}} \biggl[\langle \sigma v\rangle_{T_\chi} - \langle \sigma v \rangle_{2,T_\chi} \biggr] + \frac{s Y_\chi}{x \tilde{H}} \frac{(Y_\chi^{\mathrm{eq}})^2}{Y_\chi^2} \left[ \frac{y_{\mathrm{eq}}}{y} \langle \sigma v \rangle_{2,T} - \langle \sigma v \rangle_T \right] + \frac{\tilde{g}}{x } \frac{\langle p^4/E^3 \rangle_{T_\chi}}{3 T_\chi},
 \end{align}
  the number density is expressed through the yield $Y_\chi = n_\chi/s$ and the temperature through the dimensionless version $y = T_\chi m_\chi s^{-2/3}$ of the second momentum moment with $y_{\mathrm{eq}} = m_\chi T s^{-2/3}$ and $\tilde{H} = H/\tilde{g}$.

The temperature subscript on the (thermal) averages indicates whether the SM or DM distribution is used to perform the average while a `2' as subscript refers to the additional appearance of $p^2/3E$ in the averaging process. For a more detailed discussion of the cBE approach as well as the precise definition of $\langle \sigma v \rangle_{2}$ and $\langle p^4/E^3 \rangle_{T_\chi}$, we refer to Refs. \cite{Binder:2017rgn,Binder:2021bmg}. Let us close this brief summary of the fluid equations by highlighting that the elastic collision term does not drop out for the second moment in contrast to the zeroth moment but enters through the average
 \begin{equation}
     \langle C_{\mathrm{el}} \rangle_2 = \frac{g_\chi}{n_\chi T_\chi} \int \frac{\dd[3]{p}}{(2\pi)^3} \frac{p^2}{3 E} C_{\mathrm{el}},
     \label{eq:Cel2}
 \end{equation}
which simplifies to $ \langle C_{\mathrm{el}} \rangle_2  \to \gamma(T)(y_{\mathrm{eq}}/y - 1)$ for the Fokker-Planck approximation in Eq.~\eqref{eq:CFP}, where we have neglected the additional relativistic $\langle p^4/E^3 \rangle_{T_\chi}$ correction from $C_{\mathrm{FP}}$.  
In all cases, today's DM relic density is determined from the DM number density $n_\chi(T_\infty)$ long after freeze-out through
\begin{equation}
\Omega_\chi h^2=\kappa\frac{n_\chi(T_\infty) m_\chi}{\rho_c/h^2}\frac{h_{\rm eff}(T_0)}{h_{\rm eff}(T_{\infty})}\left(\frac{T_0}{T_{\infty}}\right)^3,    
\end{equation}
with $\kappa=1(2)$ if $\chi$ is self-conjugate (non-self-conjugate), today's photon temperature $T_0$, the photon temperature $T_{\infty}$ long after freeze-out and the critical density $\rho_c$.

\subsection{Discretization technique and numerical strategy}
To solve Eq.~\eqref{eq:boltzFinal} for the phase space distribution numerically, we restrict the comoving momenta to lie in the range $10^{-2}\leq q\leq 10^2$ and discretize it into $N=200$ points $q_1$, \dots, $q_{N}$ since we found that this range together with the number of momentum slices allows to accurately solve the full Boltzmann equation for a wide range of DM masses. 
Inspired by \texttt{CLASS}~\cite{Blas:2011rf} and \texttt{FortEPiaNO}~\cite{Gariazzo:2019gyi}, we choose to discretize the comoving momentum space according to the Gauss–Laguerre (GL) quadrature formula, a method designed for integrals of the type
\begin{equation}
    \int_0^\infty x^\alpha e^{-b x} f(x) \dd{x} \approx \sum_{i=0}^{N-1} w^{(\alpha,b)}_i f(x_i).
\end{equation}
This quadrature rule is also valid for the two-dimensional energy or momentum integrals appearing in the collision term since the integrand is, through the phase space distribution functions, exponentially suppressed in both integration variables.
A suitable, yet arbitrary, choice is to generate the weights $w^{(\alpha,b)}_i$ for the parameters $\alpha=0$ and $b=1/2$ which is e.g. possible with the GNU scientific library. As the distribution functions we encounter are not exactly proportional to $e^{-b x}$, we define $e^{b x}$ into $f(x)$. 

If, on the other hand, the scattering term is approximated by the Fokker-Planck type operator in Eq. \eqref{eq:CFP}, we deviate from the GL prescription and choose a logarithmic spacing instead as we use sixth order central difference formulas for the numerical evaluation of the first and second momentum derivatives to achieve a high accuracy, i.e. we take $f_\chi$ as a function of $\log q$ with
a uniform spacing $\Delta\!\ln q = \ln(q_{i+1}/q_i)$. 
For the in total six points outside the solution domain, the conditions $f_{-2}=f_{-1}=f_0=f_1$ as well as  $f_{N}=f_{N+1}=f_{N+2}=f_{N+3}$ with $f_l = f_\chi(q_l,t)$ are used. In addition, the momentum derivatives are computed in log-space, $\ln f_\chi(p,t)$, to obtain also a high accuracy for large momenta where the distribution function can differ by  several orders of magnitude for neighboring points.

Even though the elastic collision term manifestly conserves the number of particles in the continuum limit, the discretized version can lead to a spurious change of the comoving number density in the initial high-temperature regime. We therefore follow the same prescription as in Ref.~\cite{Binder:2021bmg} and assume kinetic equilibrium if the ratio $\gamma(T)/H(T)$ is larger than $10^5$ and solve the nBE instead.

The set of $N$ Boltzmann equations obtained after discretizing Eq.~\eqref{eq:boltzFinal} are stiff differential equations which require special integration routines to overcome the stiffness difficulty. For this purpose, the \texttt{CVODE} solver based on the backward differentiation formula of the \texttt{SUNDIALS} library \cite{hindmarsh2005sundials,gardner2022sundials} is used in our \texttt{C}-code.

\section{Leptophilic forbidden dark matter}
\label{sec:forbiddenDM}
The forbidden DM model under consideration consists of a Dirac fermion $\chi$ as DM with $g_\chi=2$ degrees of freedom, which couples only to SM leptons through a real scalar $\phi$ as mediator. After electroweak symmetry breaking, the \emph{effective} Lagrangian reads
\begin{equation}
        \mathcal{L} = \mathcal{L}_{\mathrm{SM}} + \frac{1}{2} \phi (\Box - m_\phi^2) \phi + \bar{\chi} (i \slashed{\partial} - m_\chi) \chi - g^S_{ij} \phi \bar{l}_i l_j - i g^P_{ij} \phi \bar{l}_i \gamma_5 l_j - g^S_\chi \phi \bar{\chi} \chi - i g_\chi^P \phi \bar{\chi} \gamma_5 \chi, 
    \label{eq:Lagrangian}
\end{equation}
with the flavor indices $i,j = e,\mu,\tau$. In the absence of lepton flavor violation, the associated DM annihilation cross section into leptons reads
\begin{equation}
    (\sigma v_{\mathrm{lab}})_{\chi\bar{\chi}\to l\bar{l}} = \frac{4\pi\sqrt{1 - 4 m_l^2/s}}{ s - 2 m^2_\chi}  \frac{\alpha^P_{\chi} s+ \alpha^S_\chi \left(s-4 m_\chi^2\right)}{(s-m_\phi^2)^2 + \Gamma_\phi^2 m_\phi^2 }    \left(\alpha^P_{ll} s+ \alpha^S_{ll} \left(s-4  m_l^2\right)\right),
    \label{annll}
\end{equation}
with the laboratory velocity $v_{\mathrm{lab}} = \left[s(s - 4 m_\chi^2)\right]^{1/2}/(s - 2m_\chi^2)$, 
$\alpha^{S,P}_{i(i)}=(g^{S,P}_{i(i)})^2/(4\pi)$
and the decay width of the mediator
\begin{equation}
    \Gamma_\phi = \frac{1 }{2 }\sum_{i =\chi,l} \sqrt{m_\phi^2-4 m_i^2} \left[\alpha^S_{i(i)}
   \left(1-\frac{4 m_i^2}{m_\phi^2}\right) + \alpha^P_{i(i)} \right],
\end{equation}
 where the sum runs over all kinematically accessible decay channels.
The phenomenology of this model in the sub-threshold regime  and its relic density within the standard kinetic equilibrium approach for a small mass difference $\delta = (m_l - m_\chi)/m_\chi$ have already been explored extensively in the work by \emph{D’Agnolo et al.} \cite{DAgnolo:2020mpt}. 
Given that Eq.~\eqref{eq:Lagrangian} seems to contradict invariance under the electroweak gauge group, it is necessary to find an ultraviolet-complete alternative. This is e.g. possible by extending the two-Higgs-doublet model by an $\operatorname{SU}(2)_L$ scalar singlet which then couples to the new fermion acting as DM \cite{Herms:2022nhd}. With a focus on the early kinetic decoupling effect, the study of the effective theory has been repeated in Ref. \cite{Liu:2023kat} where the relic density was computed with the cBE treatment based on the small momentum transfer approximation resulting in a reduction of the viable parameter space. However, as already pointed out by Refs.~\cite{Binder:2021bmg,Liu:2023kat}, the FP approximation breaks down if the particles scattering off of each other are very close in mass, as it is the case in forbidden scenarios. For this reason, the main objective of this work is to include the full elastic collision term not only in the set of coupled Boltzmann equations but also to go beyond the cBE treatment and investigate the relic density as a solution of the full Boltzmann equation at the level of the phase space density (fBE).

Similar to the original work~\cite{DAgnolo:2020mpt}, we identify throughout this whole paper the DM couplings with the values $\alpha_\chi^S=0$ and $\alpha_\chi^P=\SI{0.1}{}$, which we define at the energy scale corresponding to $m_\phi$. The scalar coupling is set to zero since expanding the squared CM energy in the laboratory velocity $s = 4 m^2_\chi (1 + v^2_{\mathrm{lab}}/4) + \order{v^4_{\mathrm{lab}}}$ shows that only the coupling $\alpha_\chi^P$ contributes (at tree-level) an $s$-wave component. In addition, this choice ensures that both couplings remain perturbative $\alpha^S_\chi,\alpha_\chi^P < 1$ below $\SI{1}{\tera\electronvolt}$ for the investigated mediator mass range $\SI{0.1}{\giga\electronvolt}\leq m_\phi\leq \SI{100}{\giga\electronvolt}$ as dictated by the one-loop renormalization group equation \cite{Balian:1976vq}
\begin{align}
    \mu\dv{\alpha_\chi^{S/P}}{\mu} = \frac{5}{2\pi} \alpha_\chi^{S/P}(\alpha_\chi^{S/P} + \alpha_\chi^{P/S}) .
\end{align}
We also do not include $\phi$-coannihilations in our calculation which is justified since we restrict our investigation to the mass region $\tilde{r} = m_\phi/m_l \geq 1.25$. We have therefore already set the cubic (quartic) $\phi^3(\phi^4)$-interaction terms explicitly to zero. We also neglect scatterings off the mediator, as the same exponential suppression of the averaged coannihilation cross section due to this mass splitting appears in the associated momentum transfer rate $\gamma_\phi(x) \sim e^{- x \tilde{r}}$. However, as this line of reasoning only holds assuming $i)$ the mediator is in equilibrium with the SM and $ii)$ there are no large hierarchies present between the DM and lepton couplings as well as the mediator self-couplings which could compensate the suppression at early times, it would be interesting to drop these assumptions and include the mediator in the network of momentum-dependent Boltzmann equations.

\subsection{The relic density beyond kinetic equilibrium}

\begin{figure}[th]
    \centering
    \includegraphics[width=.495\textwidth]{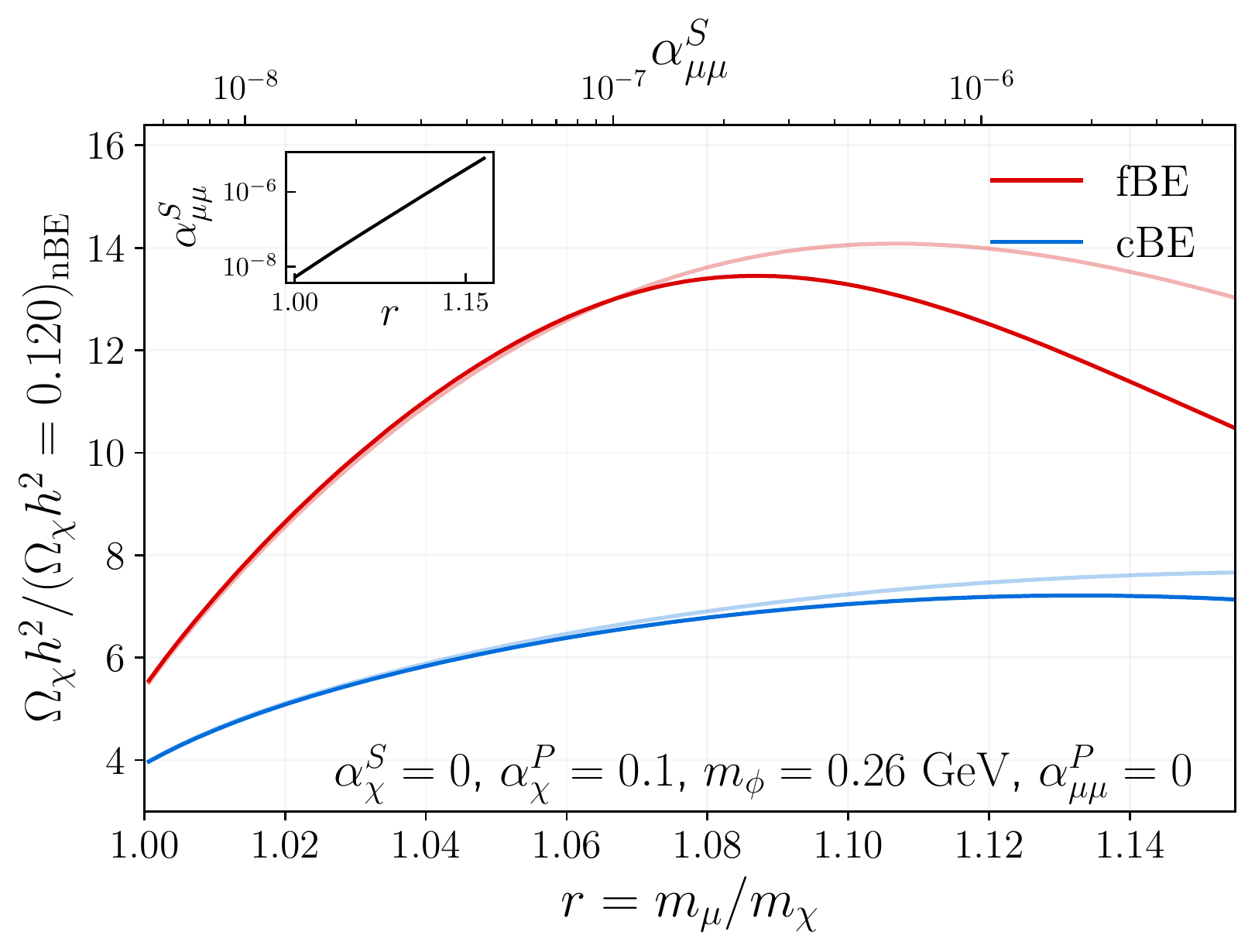}
    \includegraphics[width=.495\textwidth]{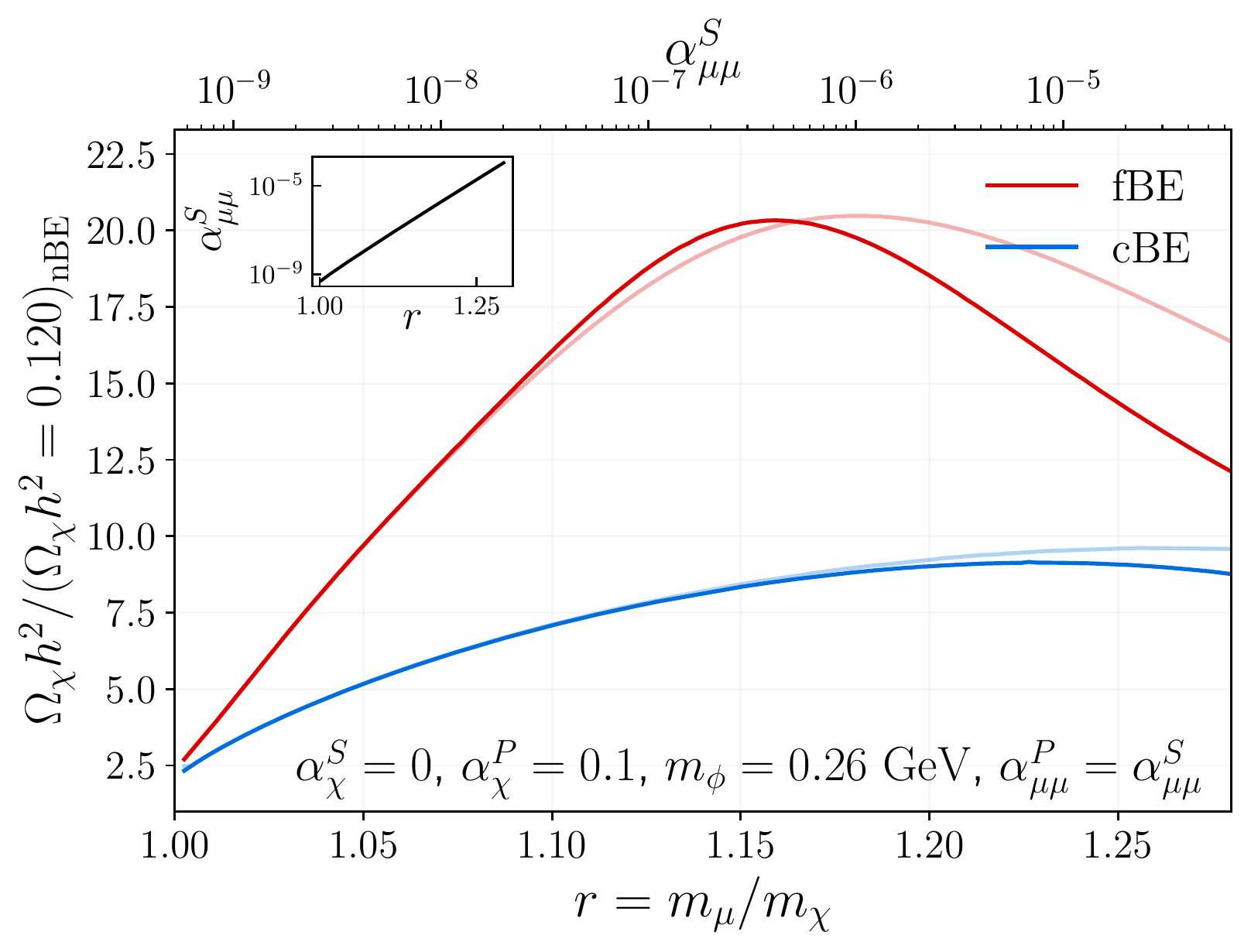}
    \includegraphics[width=.495\textwidth]{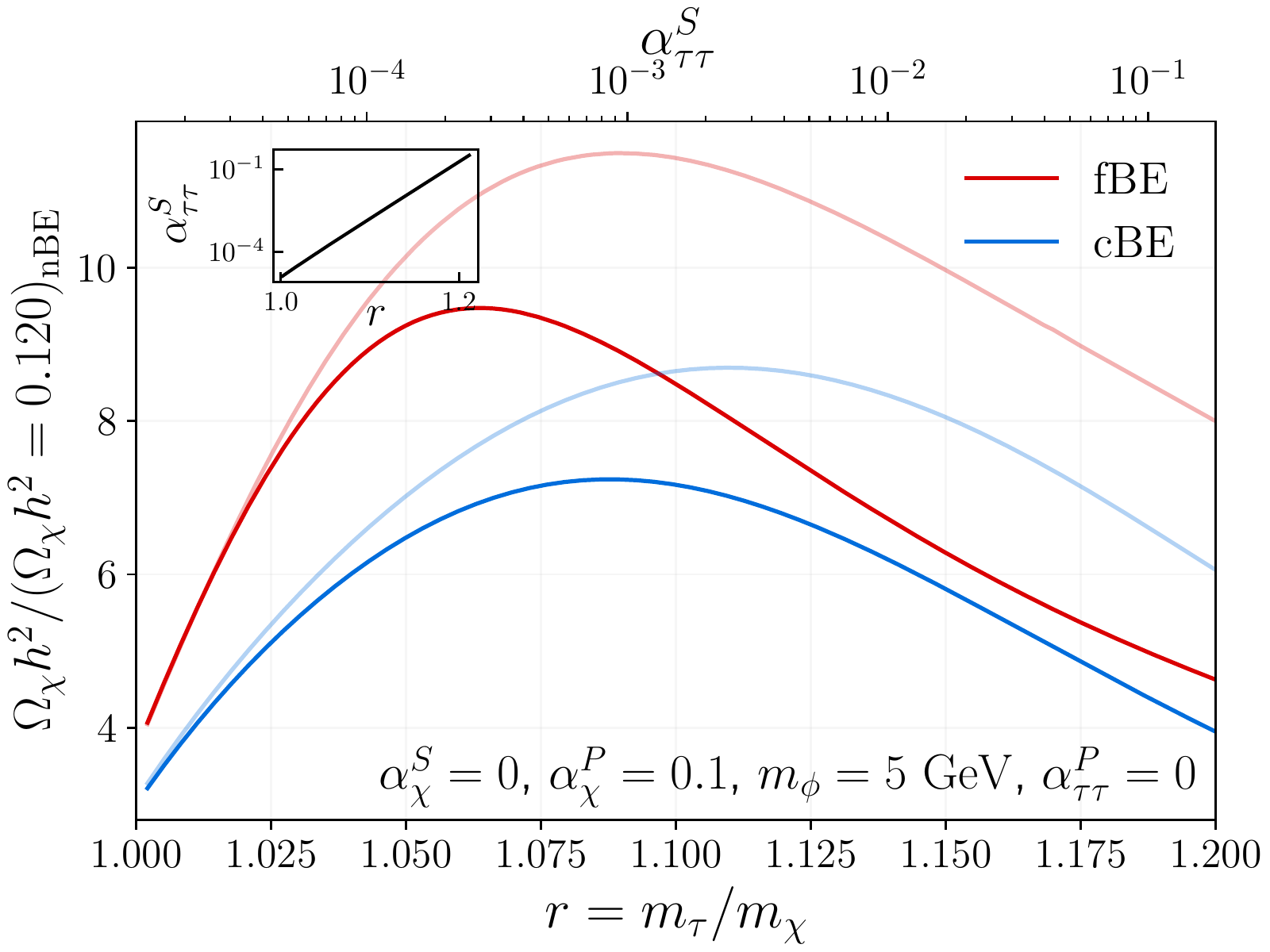}
    \includegraphics[width=.495\textwidth]{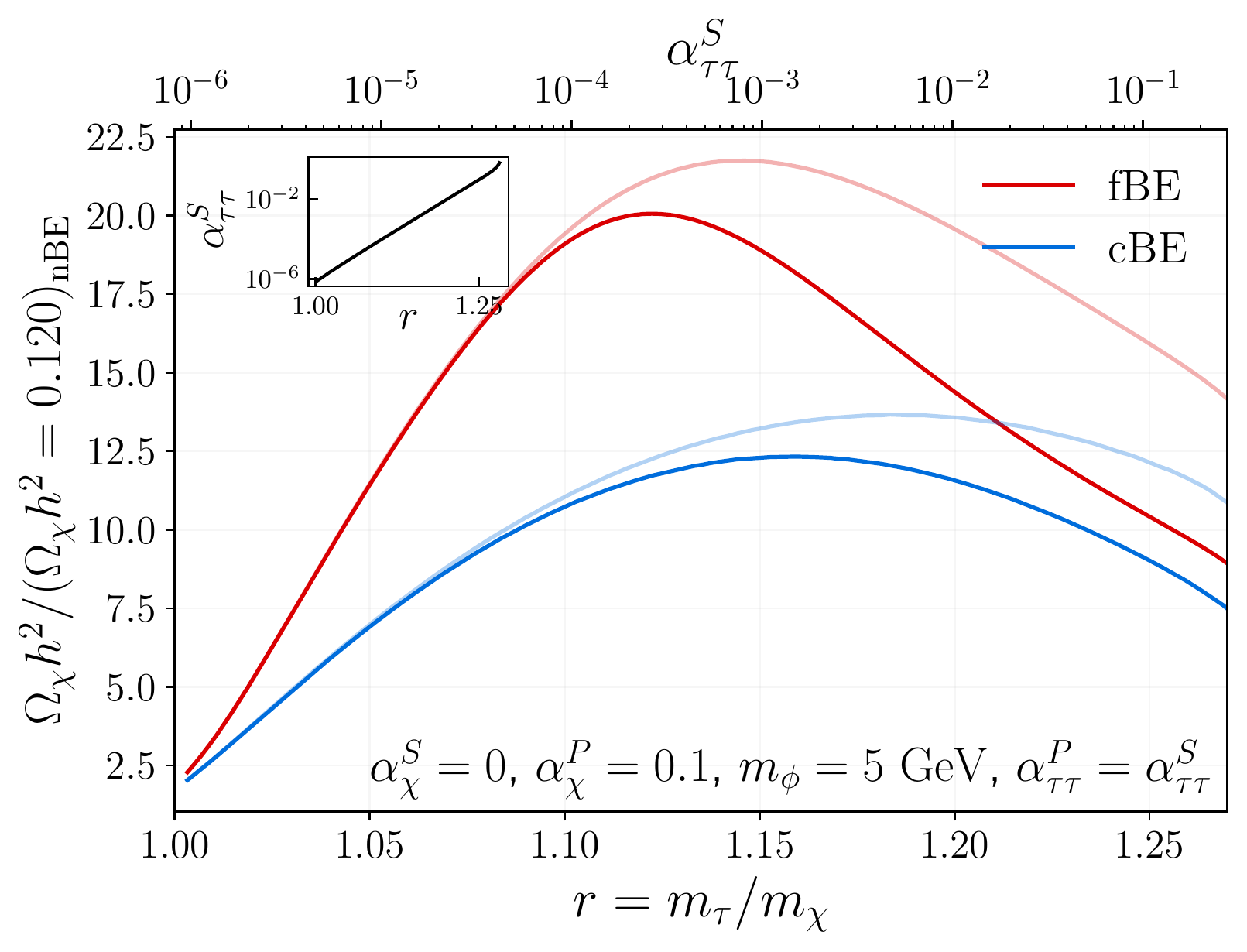}
     \caption{Ratio of the relic densities obtained with the cBE (blue) as well as the fBE (red) treatment and the number density approach as a function of the mass ratio $r=m_l/m_\chi$ for annihilations into muons (upper row) and tau leptons (lower row). In the left column, the pseudoscalar coupling is turned off whereas it is identical to the scalar coupling on the right. In all cases, the lepton coupling is chosen such that the result from the nBE method matches the experimentally observed relic abundance which imposes, to a very good approximation, an exponential relation between coupling and $r$ (figure insets). The lines corresponding to the full elastic collision term are non-transparent, whereas the associated results obtained with the Fokker-Planck approximation are displayed with a lighter opacity. }
    \label{fig:Oh2Ratio}
\end{figure}

\begin{figure}[t]
    \centering
    \includegraphics[width=.495\textwidth]{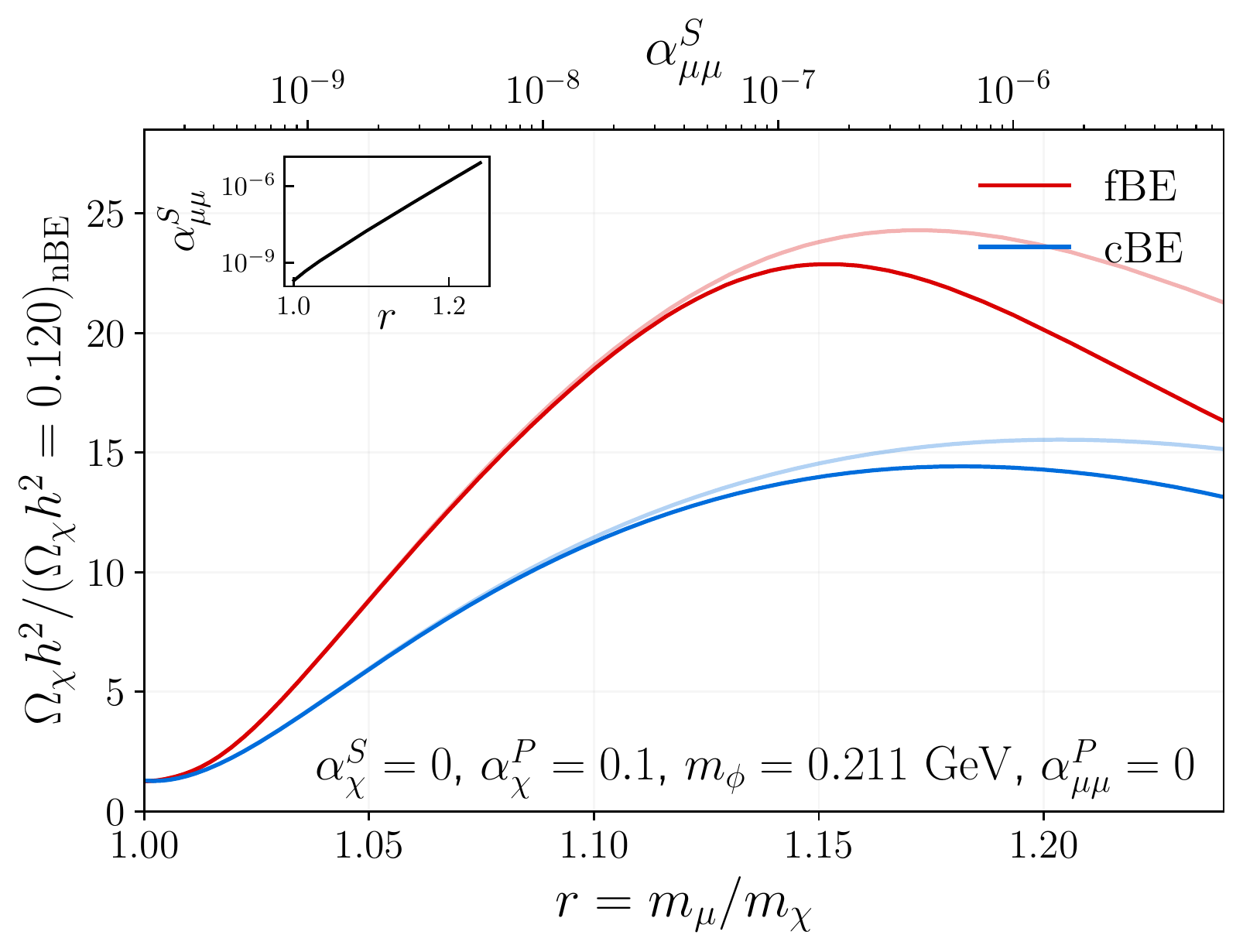}
    \includegraphics[width=.495\textwidth]{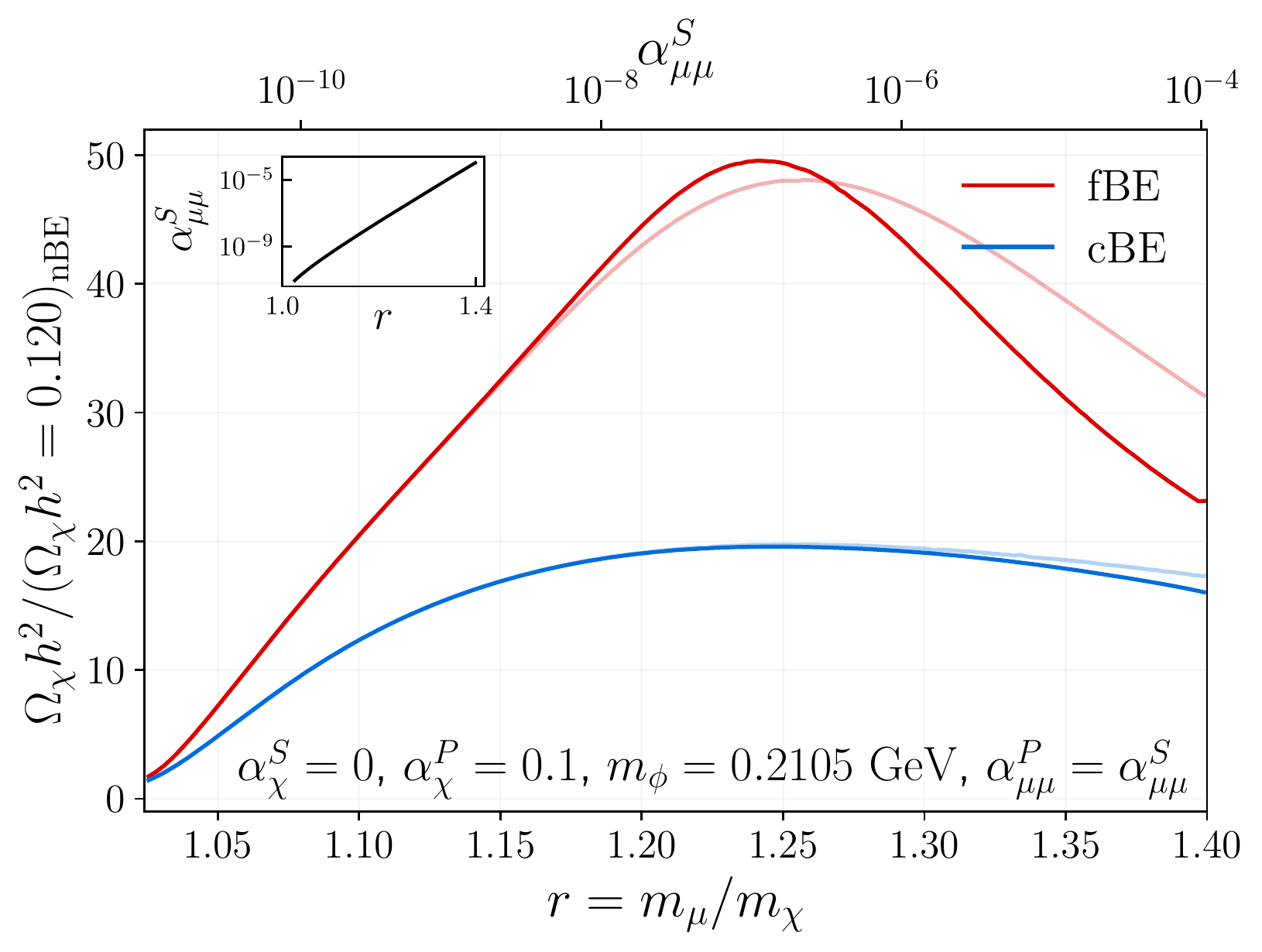}
    \includegraphics[width=.495\textwidth]{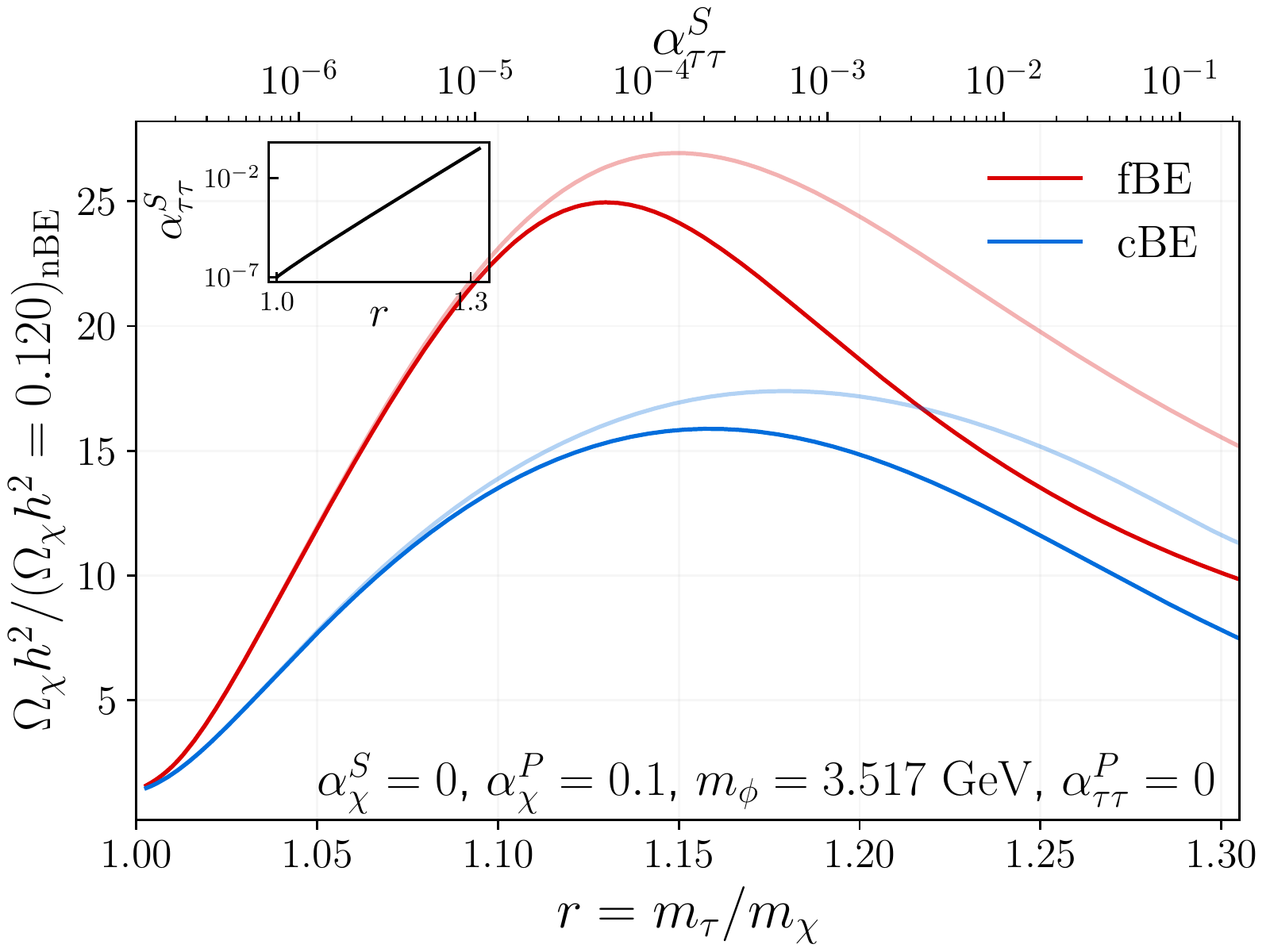}
    \includegraphics[width=.495\textwidth]{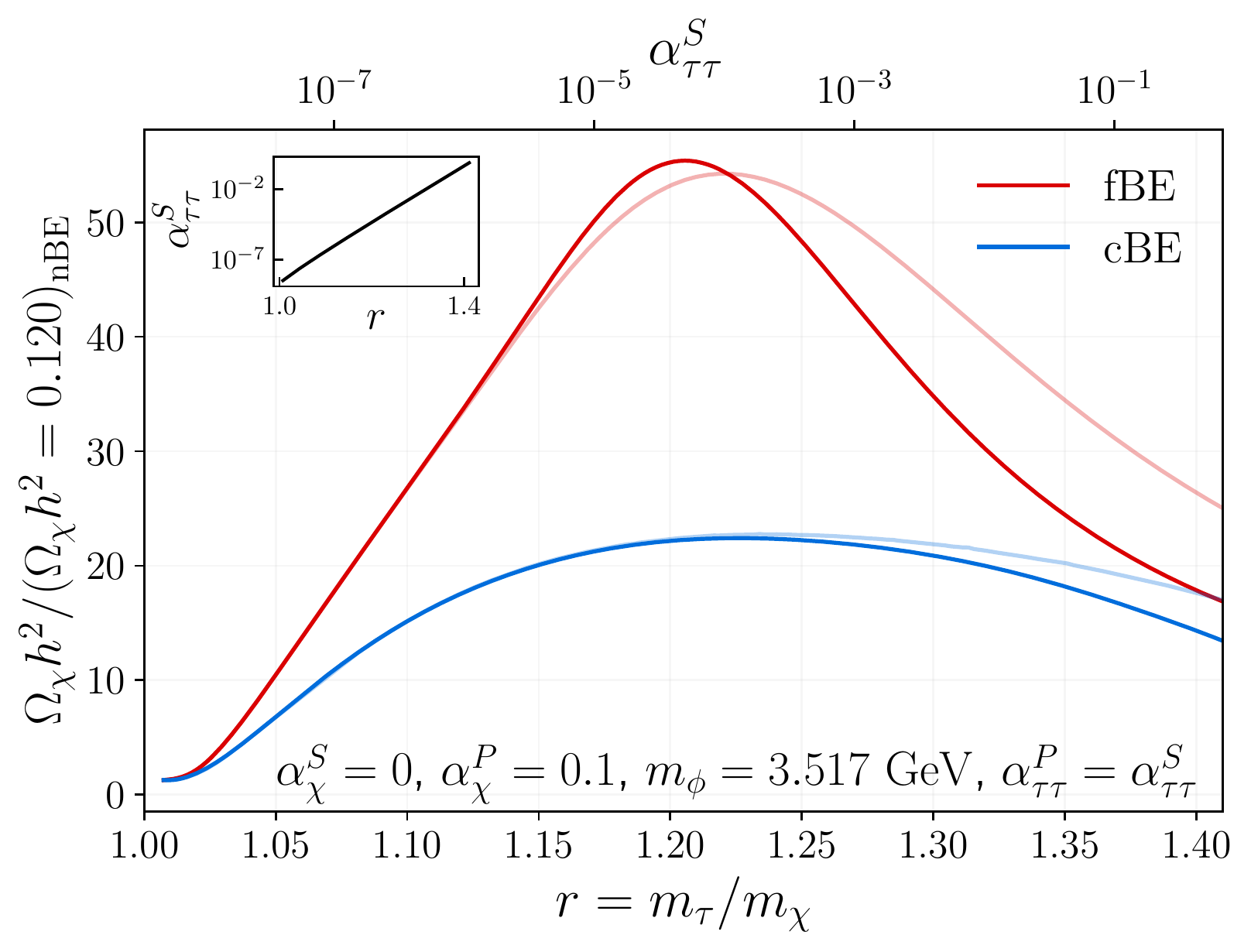}
    \caption{Same as Fig. \ref{fig:Oh2Ratio}, but for a $m_\phi$ that roughly corresponds to twice the muon (tau) mass in the upper (lower) panel. }
    \label{fig:Oh2RatioRes}
\end{figure}

\begin{figure}[t]
    \centering
    \includegraphics[width=0.495\textwidth]{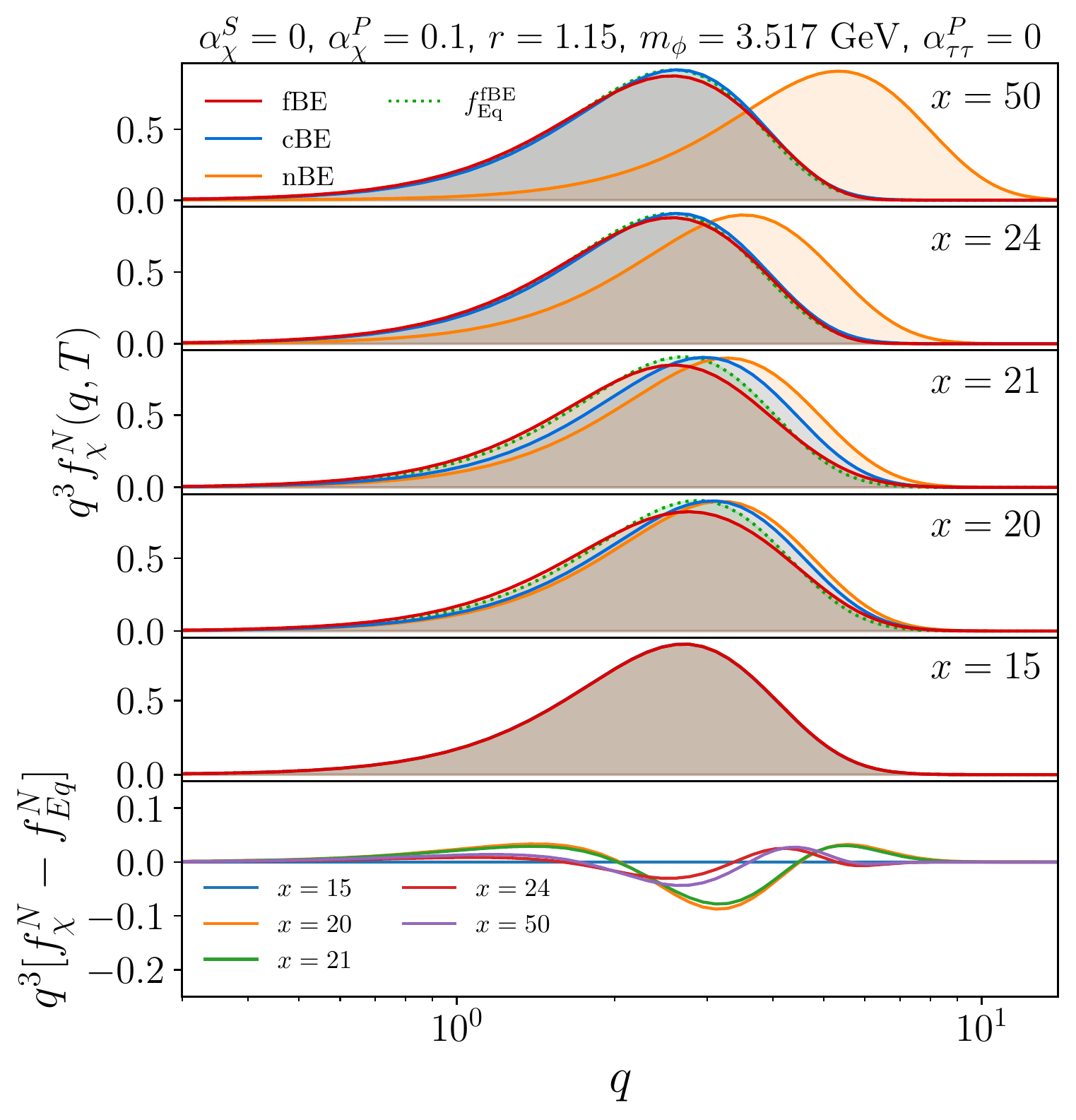}
    \includegraphics[width=0.495\textwidth]{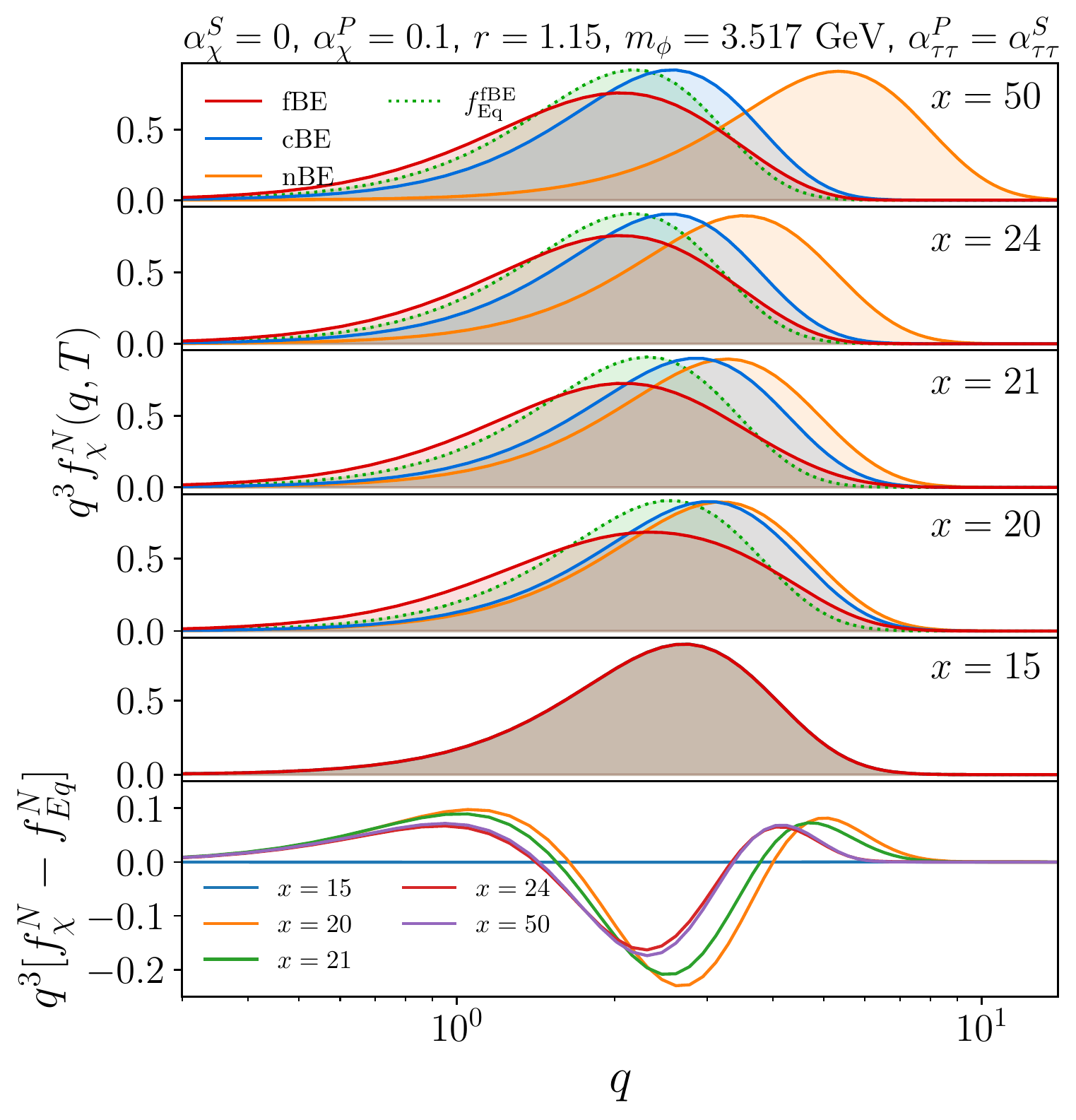}
     \caption{Upper panels: Snapshots of the normalized dark matter phase space distributions $f^N_\chi$ taken at different values of $x$ plotted against the comoving momentum $q$ for the case of annihilation into tau leptons with a mediator mass close to the di-tau resonance. The distributions from the cBE and fBE are both obtained with the full collision term. Lower panels: The difference of the phase space distribution function between the fBE and the corresponding equilibrium distribution evaluated at the DM temperature $T_\chi$ as defined in Eq.~\eqref{eq:TChiPS}. }
    \label{fig:PSBPtauRes}
\end{figure}

We begin the numerical analysis by comparing the DM relic density obtained from solving the Boltzmann equations (cBE and fBE) using the FP approximation on one hand and the full collision term on the other. 

In order to make a direct comparison with Ref.~\cite{Liu:2023kat} possible, our results for the four benchmarks considered in Ref.~\cite{Liu:2023kat} are displayed in Fig.~\ref{fig:Oh2Ratio}. 
The relic density obtained with the cBE and fBE approaches for the two different implementations of the elastic collision term relative to the number density result are shown in Fig.~\ref{fig:Oh2Ratio} as a function of the inverse DM mass. For every $m_\chi$, the lepton coupling $\alpha^S_{ll}$ is fixed by the requirement that the nBE result matches the experimentally observed relic density.  Curves in bold colors correspond to the use of the full collision term, whereas curves with a lighter hue correspond to the use of the FP approximation. The upper panels deal with annihilation into muons, while the lower ones display the results for tau leptons in the final state. Furthermore, the left panels shows the results for $\alpha^P_{ll}=0$, while the right panels are for $\alpha^S_{ll}=\alpha^P_{ll}$. We find in this model that the early kinetic decoupling effect increases the relic density significantly and that the small-momentum approximation in general implies a smaller contribution from the elastic scattering collision term, which then leads to an overestimation of the relic density. Put differently, the full collision term keeps the DM distribution closer to equilibrium and, as a consequence, moves the associated relic density towards the number density result. This can be clearly seen in Fig.~\ref{fig:Oh2Ratio} by comparing darker and lighter hues of same-color curves. It is worth mentioning that the FP approximation tracks the full collision solution for a significant range of values of $r$ in most of the cases. However, the two solutions significantly depart from each other for larger $r$ values with the full collision solution in the fBE case approaching the nBE solution. For $r=1$, DM still does not maintain kinetic equilibrium for this particular benchmark since for decreasing $r$, $2 m_\chi$ moves towards the mediator mass and therefore the distribution function starts to be affected by the resonance. 

\begin{figure}[t]
    \centering
    \includegraphics[width=0.495\textwidth]{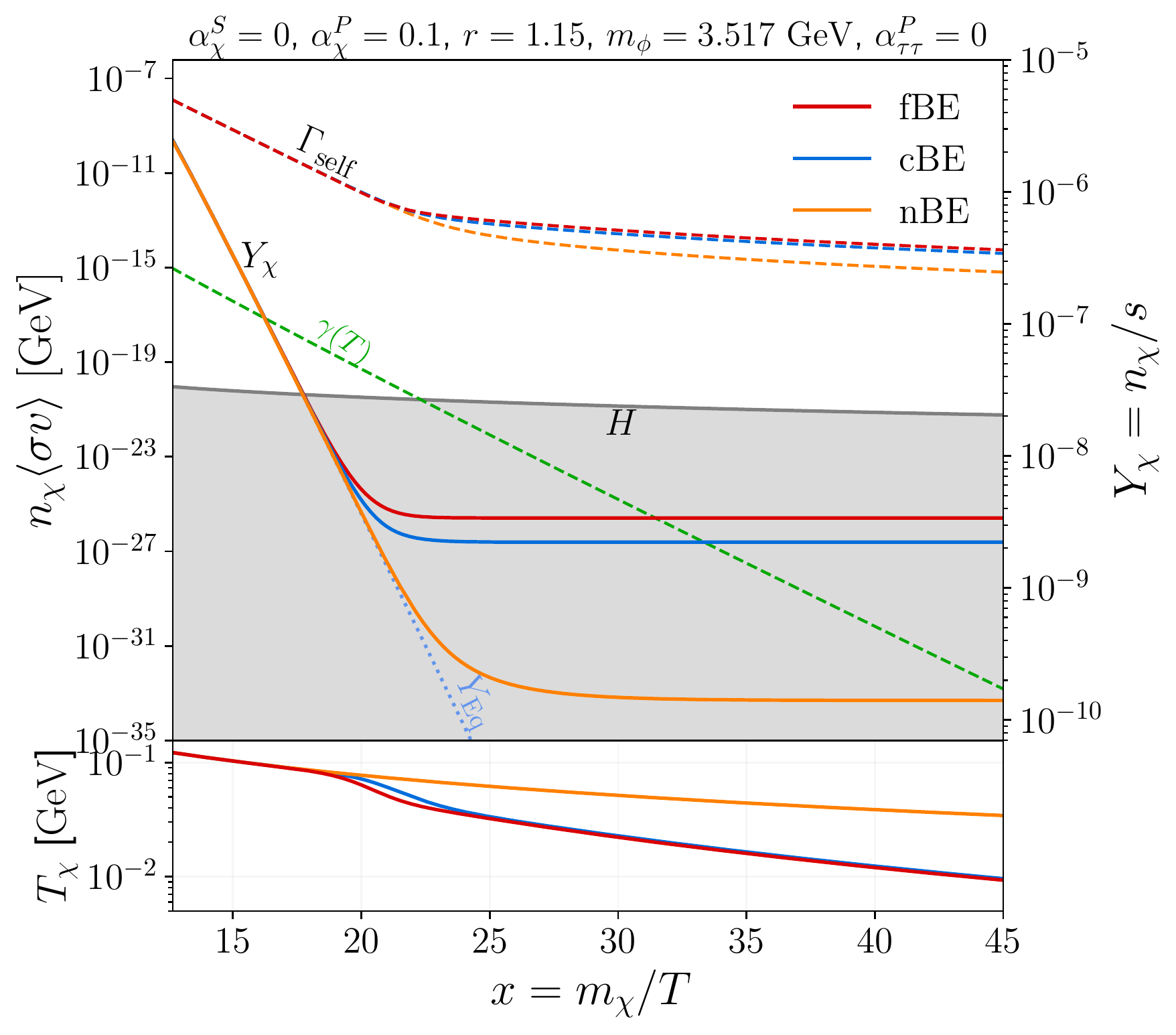}
   \includegraphics[width=0.495\textwidth]{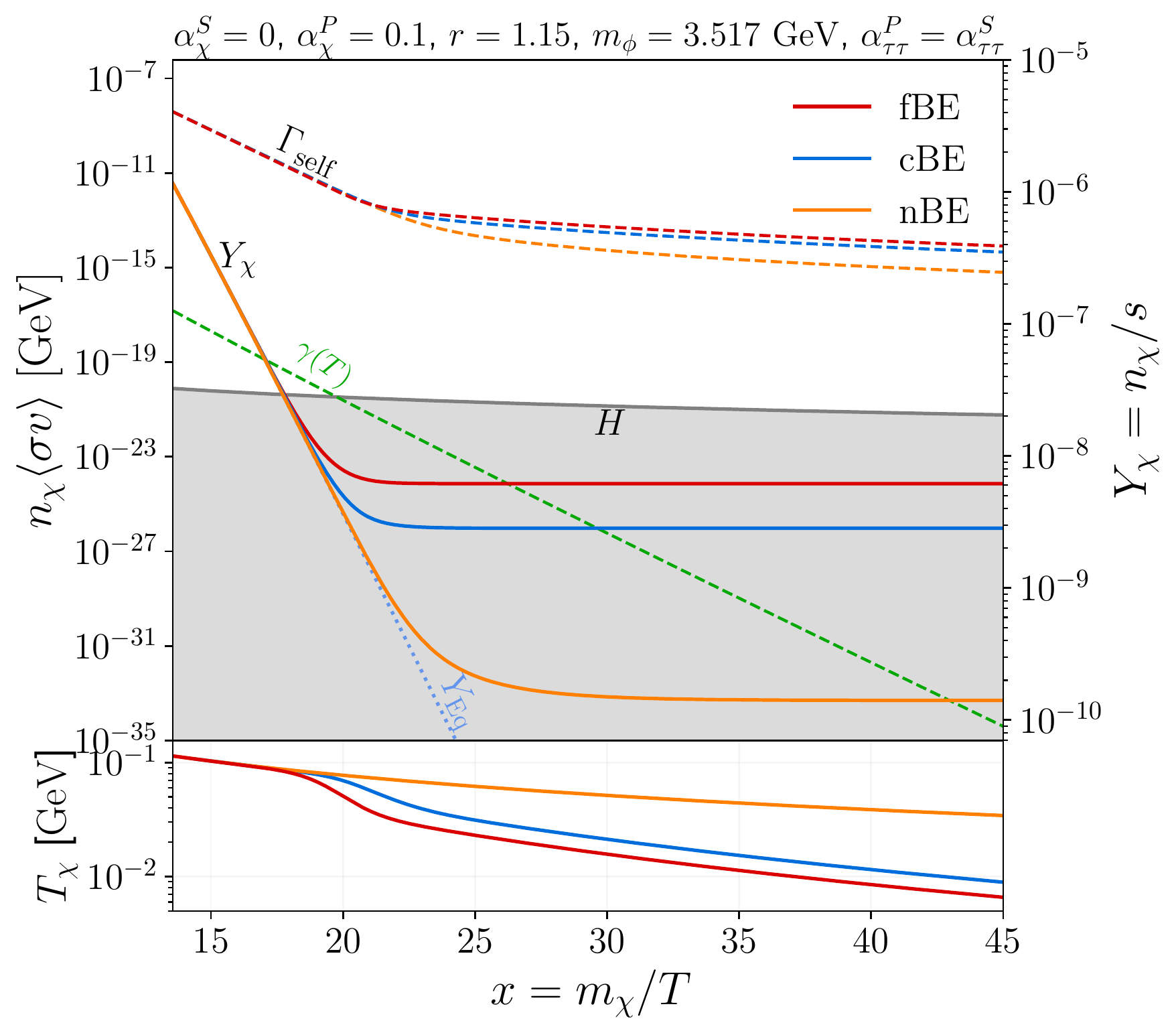}
    \caption{Evolution of the self-scattering rate $\Gamma_{\mathrm{self}}$, the yield parameter $Y_\chi$, the momentum exchange rate $\gamma(T)$ and the DM temperature $T_\chi$ in $x$ as obtained with the nBE, cBE and fBE approaches for annihilations into tau leptons with $m_\phi \approx 2 m_\tau$. Note that the yield parameter $Y_\chi$ only accounts for the number of DM particles and not antiparticles and that $\Gamma_{\mathrm{self}}$ is obtained for the fBE case from the actual phase space distribution as defined in Eq.~\eqref{eq:GammaSelf}.}
    \label{fig:GSelfBPtauRes}
\end{figure}

In order to not only restrict our discussion to mediator masses far away from the di-muon or di-tau resonance, we display in Fig.~\ref{fig:Oh2RatioRes} the same quantities but with a mediator corresponding to approximately twice the muon (upper panels) or tau mass (lower panels). As a result, the DM distribution function is in this case not only driven out of equilibrium through the forbidden nature of the model, but also significantly through the resonance. 
This enhancement can be clearly seen in the increased deviation from the nBE approach compared to the previous case where the mediator mass is much larger than the corresponding lepton mass. For example, the relic density from the fBE for the benchmark with a non-zero pseudoscalar coupling is more than 50 times that predicted using the nBE and $\gtrsim 2.5$ times than the cBE result. It should be noted that the ratio of the relic densities in Figs.~\ref{fig:Oh2Ratio} and \ref{fig:Oh2RatioRes} is larger for the cases when $\alpha_{ll}^S=\alpha_{ll}^P$ than it is when $\alpha_{ll}^P=0$. By examining Eq.~\eqref{annll}, we see that this is due to the fact that the $s$-wave component of the annihilation cross-section is suppressed for $\alpha_{ll}^P=0$ through the small mass difference between leptons and DM which is not the case for $\alpha_{ll}^S=\alpha_{ll}^P$. On a qualitative level, this effect can also be understood through the difference between the phase space distribution functions obtained with the fBE and the corresponding equilibrium distribution functions $f_{\rm Eq}^{\rm fBE} \sim e^{-E/T_\chi}$ which is shown in the lowermost panels of Fig.~\ref{fig:PSBPtauRes} for five values of $x$ for both choices of $\alpha^P_{\tau\tau}$ and $r=1.15$. Here, the DM ``temperature'' $T_\chi$ is computed from $f_\chi$ itself as defined in Eq.~\eqref{eq:TChiPS}. It becomes clear that the strong velocity dependence of the annihilation cross section leads to dips in the distribution function which cause the departure from equilibrium and are more pronounced for the $\alpha_{ll}^S=\alpha_{ll}^P$ case than for $\alpha_{ll}^P=0$ and in particular near freeze-out at $x\sim 20$. The deeper the dip the more inefficient the DM annihilation becomes due to the smaller occupation number at the relevant momenta. This correlation then leads to a higher DM relic density as can be seen in Figs. \ref{fig:Oh2Ratio} and \ref{fig:Oh2RatioRes}. From the upper panels of Fig. \ref{fig:PSBPtauRes} showing the evolution of the distribution function for five different values of $x$ it becomes clear that the distributions obtained from the cBE and the fBE evolve in general to lower momenta compared to the equilibrium distribution evaluated at the photon temperature as the high momentum DM particles get depleted in order to overcome the annihilation threshold. It is also noticeable that the solution of the fBE departs away from $f_{\rm Eq}^{\rm fBE}$ at $x\sim 20$. While this deviation remains strong for the $\alpha^P_{\tau\tau}=\alpha^S_{\tau\tau}$ case shown in the right panel, the deviation almost vanishes at later times for the $\alpha^P_{\tau\tau}=0$ case in the left panel.
 This difference can be understood by comparing the momentum transfer rate $\gamma(T)$ with the Hubble rate $H(T)$. Both are displayed in Fig.~\ref{fig:GSelfBPtauRes} as a function of $x$ with the left panel corresponding to the case $\alpha^P_{\tau\tau}=0$ while the right panel shows the case $\alpha^P_{\tau\tau}=\alpha^S_{\tau\tau}$. It is clear that for $\alpha^P_{\tau\tau}=0$, $\gamma(T)$ is still comparable to $H(T)$ around $x\sim 24$ which means that elastic scattering is still effective enough to keep the phase space distributions from deviating too far away from equilibrium as we have seen in the left panel of Fig.~\ref{fig:PSBPtauRes}. However, for $\alpha^P_{\tau\tau}=\alpha^S_{\tau\tau}$, $\gamma(T)$ is already much smaller than $H(T)$ at $x\sim 24$ which suggests that elastic scatterings are not strong enough leading to a larger deviation away from equilibrium. Fig.~\ref{fig:GSelfBPtauRes} also shows the evolution of the DM yield in the nBE, cBE and fBE approaches. Notice how for cBE and fBE, DM freeze-out happens earlier than for the nBE case leading to a higher relic density. The effect of the phase space distribution shifting to lower momenta can also be seen in the lower panel of Fig.~\ref{fig:GSelfBPtauRes} where we plot the DM temperature $T_\chi$. A clear drop away from the photon temperature is visible just before $x\sim 20$, where DM starts cooling faster than the SM bath. Note that the splitting between the cBE and fBE predictions can be attributed to the phase space distributions shown in Fig.~\ref{fig:PSBPtauRes}.

Lastly, the question remains whether for this DM model with the particular choice $\alpha^S_\chi = 0$ and $\alpha^P_\chi = 0.1$ for the DM couplings, the cBE or fBE approach gives a more correct description of the freeze-out process. Since the cBE framework becomes exact under the assumption of maximally efficient DM self-interactions, the associated rate given by
\begin{multline}
    \Gamma_{\mathrm{self}} = 2 n_\chi \langle \sigma_{\mathrm{self}} v \rangle = 2 n_\chi \frac{\int \dd[3]{p_a} \int \dd[3]{p_b}\sigma_{\mathrm{self}}  v_{\text{M\o l}} f_\chi(p_a) f_\chi(p_b)}{\int \dd[3]{p_a} \int \dd[3]{p_b} f_\chi(p_a) f_\chi(p_b)} \\
    = \frac{g^2_\chi}{(2\pi)^4 n_\chi}  \int^\infty_{m_\chi} \dd{E_a} f_\chi(p_a) \int^\infty_{m_\chi} \dd{E_b} f_\chi(p_b) \int^{s_+}_{s_-} \dd{s} \sqrt{s(s - 4 m^2_\chi)} \sigma_{\mathrm{self}}(s),
    \label{eq:GammaSelf}
\end{multline}
is shown in Fig.~\ref{fig:GSelfBPtauRes} for the same tau benchmarks used for the illustration of the evolution of the distribution functions. In Eq.~\eqref{eq:GammaSelf}, we have the integration limits $s_{\pm} = (E_a + E_b)^2 - (p_a \mp p_b)^2$, the self-scattering cross section $\sigma_{\mathrm{self}} = \sigma_{\chi\bar{\chi}\to \chi\bar{\chi}} + \sigma_{\chi\chi\to \chi\chi}$ which is for reference explicitly given in App. \ref{app:XSec} and the factor two in front of the number density accounting for particles and antiparticles. For the nBE and cBE approaches, the average in the self-interaction rate reduces to the single integral over the collision energy defined in Eq.~\eqref{eq:vsGG}. It is clear from Fig.~\ref{fig:GSelfBPtauRes} that even long after freeze-out, the self-scattering rate remains more than five orders of magnitude above the Hubble rate meaning that the cBE treatment gives in this case a more correct depiction of the DM thermodynamics. In fact, the self-interactions are so strong, that the addition of the self-scattering collision term $C_{\mathrm{self}}$ to the right-hand side of the momentum-dependent Boltzmann equation makes a numerical solution of the fBE impossible while using the same number of momentum bins, not compromising on accuracy and ensuring that the implementation of $C_{\mathrm{self}}$ conserves the number of particles. Since the self-interactions are so strong, we have also checked that these are not in conflict with astrophysical bounds~\cite{Kahlhoefer:2013dca}.

\subsection{Exclusion limits \label{sec:exclusionLimits}}

\begin{figure}[t]
    \centering
    \includegraphics[width=0.99\textwidth]{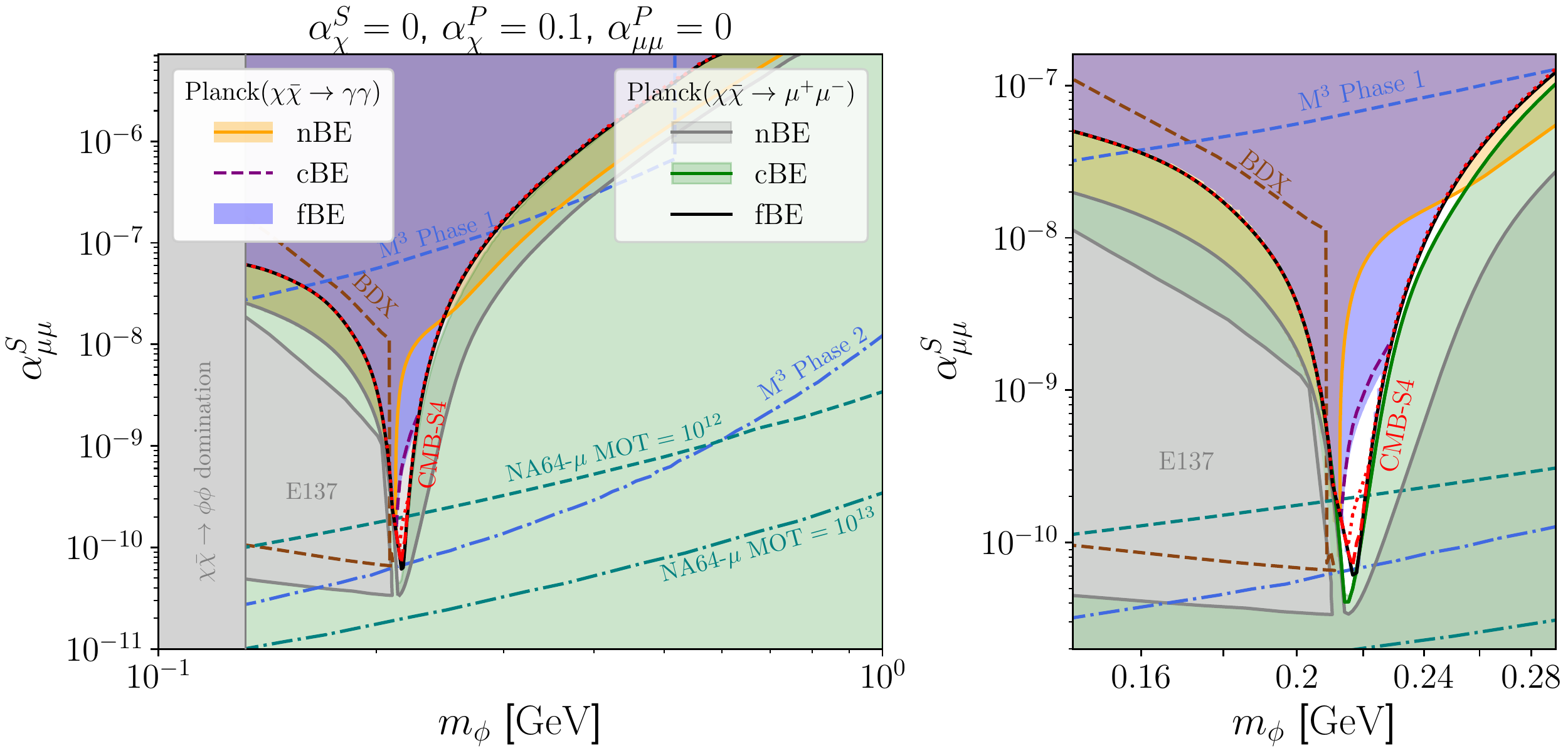}
    \includegraphics[width=0.99\textwidth]{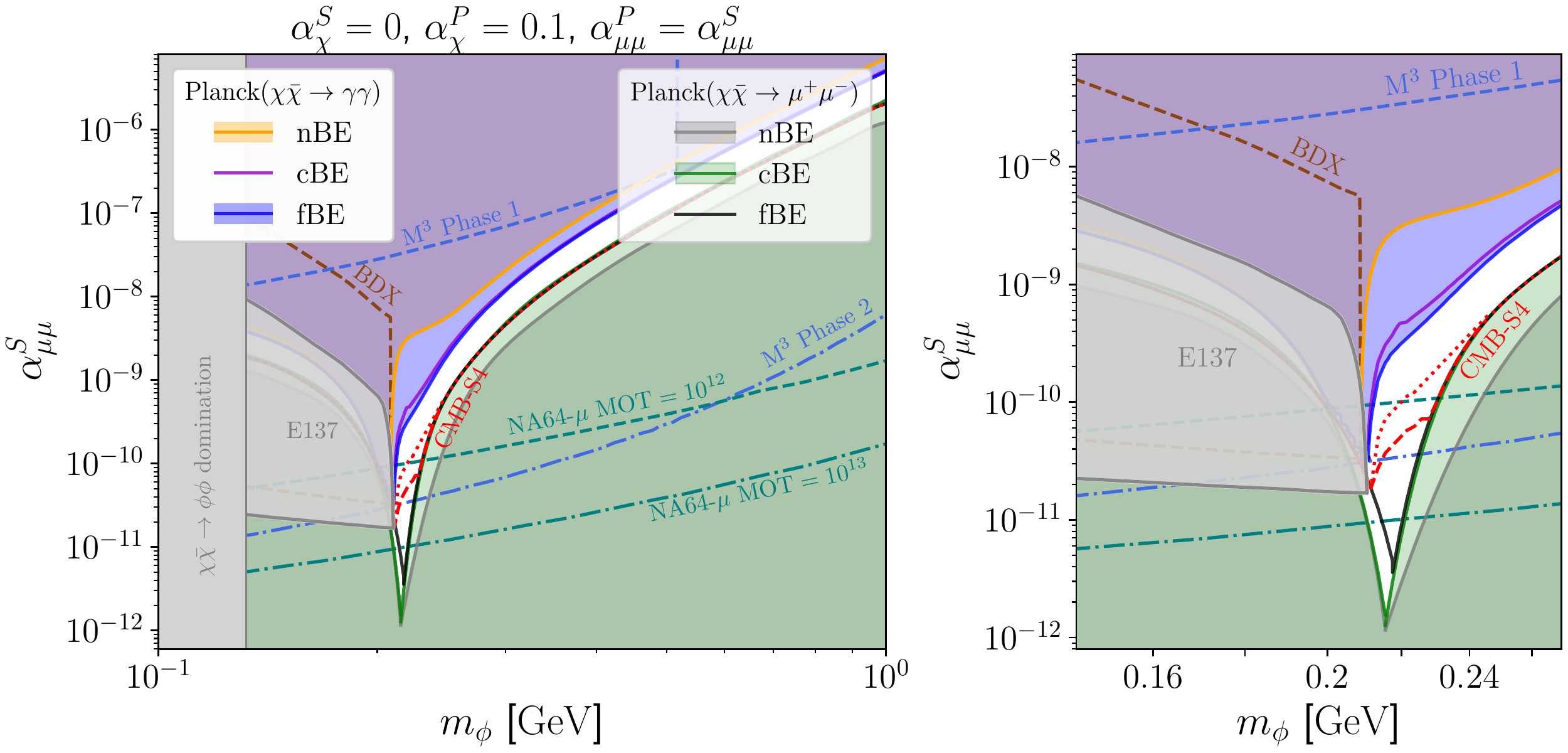}
    \caption{Exclusion limits on forbidden DM annihilations into muons in the plane spanned by the mediator mass and the scalar lepton coupling for a vanishing pseudoscalar coupling in the upper panel and for $\alpha^P_{\mu\mu}=\alpha^S_{\mu\mu}$ in the lower one. 
    For both cases, the region around the resonance $m_\phi \approx 2 m_\mu$ is shown enlarged on the right. 
    The gray, green and black lines indicate the boundary where $m_\chi$ has to equal the muon mass in order to satisfy the relic density constraint based on the nBE, cBE and fBE calculations, respectively. The region below is then excluded by CMB limits on direct annihilations into muons. At every point in the plane above this boundary the DM mass is fixed through the requirement that the DM relic density lies within the observed range where the theory value is obtained using all three computational approaches. Every approach yields a different DM mass for the same parameter point, giving three different regions excluded by CMB constraints on annihilations into photons which are shown in orange, violet and blue for the nBE, cBE and fBE approach, respectively. Displayed are also existing and projected limits from CMB-S4 and the beam-dump experiments E137, BDX, $\mathrm{M}^3$ and NA64-$\mu$. The white space corresponds to the viable region. For more details, see the main text. }
    \label{fig:BPmuScan}
\end{figure}

\begin{figure}[t]
    \centering
    \includegraphics[width=0.99\textwidth]{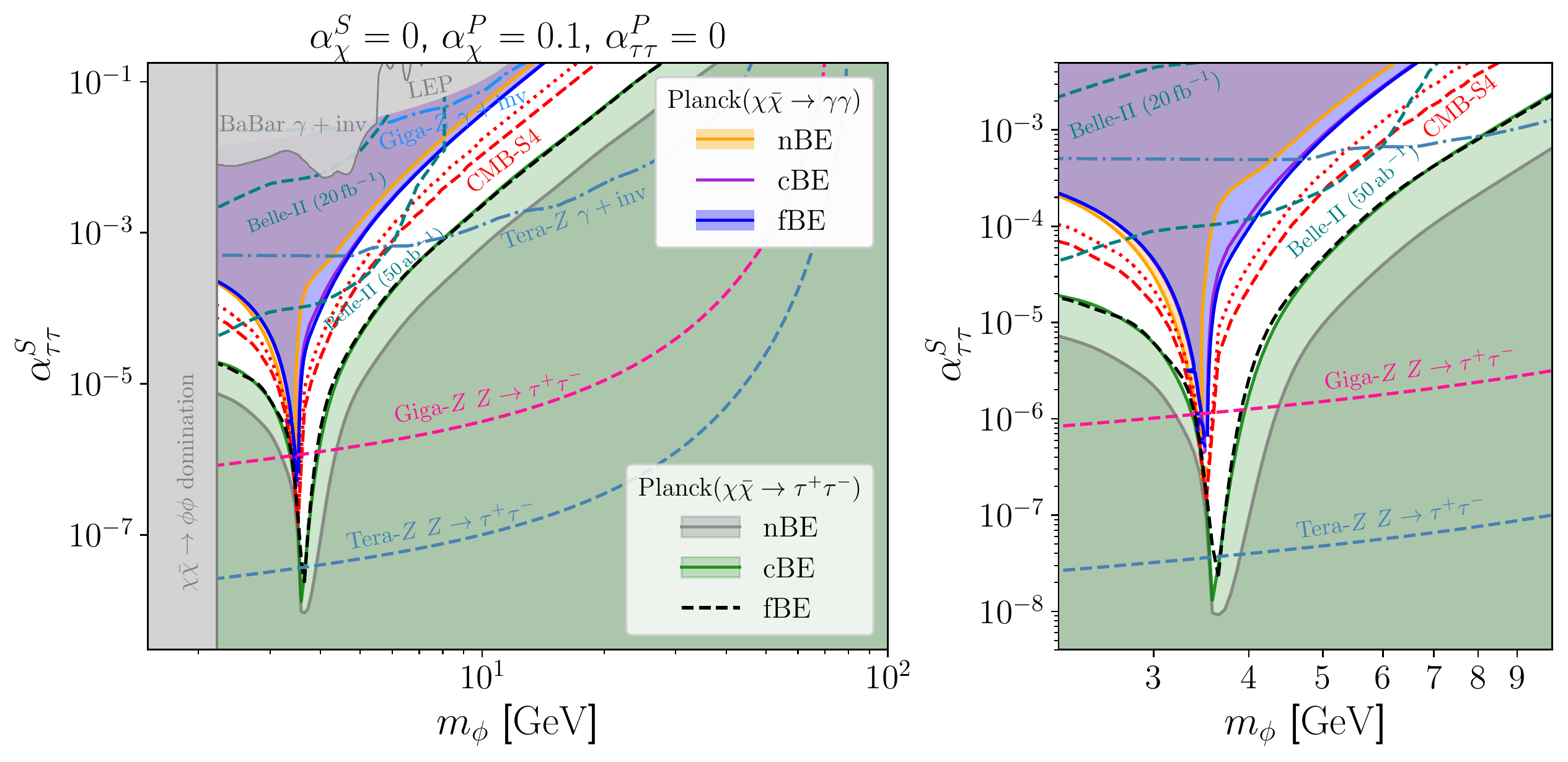}
    \includegraphics[width=0.99\textwidth]{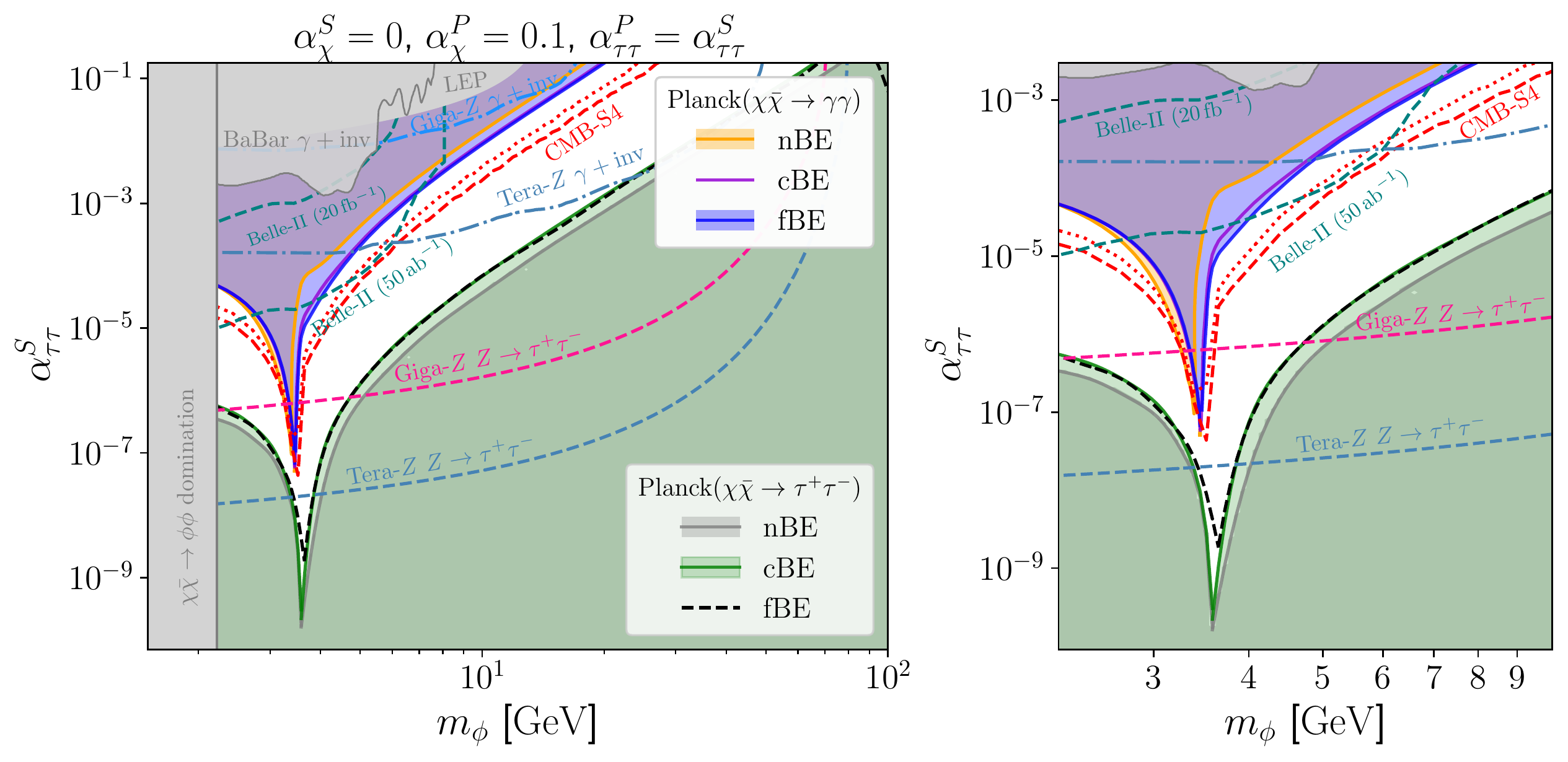}
    \caption{Exclusion limits on forbidden DM annihilations into tau leptons with the same structure as well as the same color coding regarding the CMB constraints on $\chi\bar{\chi} \to \tau\bar{\tau}$ and $\chi\bar{\chi} \to \gamma\gamma$ as in Fig. \ref{fig:BPmuScan}. In addition, existing and projected limits from LEP, BaBar, Belle-II, CMB-S4 and future $e^+e^-$-colliders with Giga-$Z$ and Tera-$Z$ options are shown. For more details, see the main text. }
    \label{fig:BPtauScan}
\end{figure}

In this section, we update the exclusion limits on forbidden DM annihilations into SM leptons based on our new calculation of the relic density. %
To compare our results with those of Ref.~\cite{DAgnolo:2020mpt}, the numerical analysis is performed in the plane spanned by the mediator mass and the scalar lepton coupling. The results for annihilations into $\mu^+\mu^-$ are shown in Fig. \ref{fig:BPmuScan} while those for annihilations into $\tau^+\tau^-$ are presented in Fig. \ref{fig:BPtauScan}. For both channels, the pseudoscalar coupling is set to zero, $\alpha_{ll}^P=0$, on the left panel and equal to the scalar coupling, $\alpha_{ll}^P=\alpha_{ll}^S$, on the right. As our analysis in the previous section shows that the small-momentum approximation still holds for significant ranges of DM masses in the forbidden regime, the calculation of the relic density in the following analysis is still based on the FP approximation with the advantage of a major reduction in run time. Even though we assume the cBE to give a more correct result, we still show for completeness the results from the fBE approach in Figs.~\ref{fig:BPmuScan} and~\ref{fig:BPtauScan}. Deriving exclusion limits based on both techniques also serves as a means to gauge the effect from DM self-scatterings.  %

For every given pair of parameters ($m_\phi$, $\alpha^S_{ll}$) in Figs.~\ref{fig:BPmuScan} and~\ref{fig:BPtauScan}, the DM mass is fixed through the requirement that the relic density corresponds to the experimentally observed value $\Omega_\chi h^2 = \SI{0.120}{}$. This calculation can be carried out efficiently using e.g. a logarithmically spaced bisection search. The thick gray, green and black curves corresponding to the nBE, cBE and fBE calculations, respectively, indicate the boundary $\delta=0$ between the forbidden and non-forbidden regions. In the region above those curves the DM mass is smaller than the corresponding lepton mass and larger below. The boundary from the cBE and fBE approaches is almost identical and both approaches leave less parameter space with $\delta>0$ compared to the nBE. The allowed white region is further constrained by terrestrial and space-based experiments as we discuss next. 
 
For both channels, the most important experimental constraint comes from DM annihilations into electromagnetically charged particles during the recombination epoch. The sensitivity of CMB anisotropies to such energy injection processes into the intergalactic medium (IGM) allows \emph{Planck} to place the upper limit $f_{\mathrm{eff}} \langle \sigma v \rangle/m_\chi  \leq \SI{3.5e-28}{\centi\meter\cubed/\second/\giga\electronvolt}$ on the annihilation parameter where the efficiency factor $f_{\mathrm{eff}}$ describes the fraction of energy that is released in the annihilation and then transferred to the IGM~\cite{Planck:2018vyg}. It should be noted that this limit is only valid for an $s$-wave dominated and therefore almost constant  annihilation cross section, i.e. $\langle \sigma v \rangle \simeq \sigma v_{\mathrm{lab}} \simeq \mathrm{const}$~\cite{Planck:2018vyg}.  
For the numerical evaluation of the efficiency factor we use the tabulated $f_{\mathrm{eff}}$ curves for DM masses below $\SI{5}{\giga\electronvolt}$ provided in Ref.~\cite{Slatyer:2015jla}. As a consequence of these robust energy injection constraints, the non-forbidden region where $\delta\leq 0$ and direct annihilations into leptons become possible is immediately ruled out. As already mentioned, the corresponding areas in Figs.~\ref{fig:BPmuScan} and \ref{fig:BPtauScan} are marked in gray for the nBE approach and in green for the cBE treatment. As the fBE result overlaps almost everywhere with the cBE one, only the boundary $\delta^{\mathrm{fBE}}=0$ is marked in black. In the forbidden region defined through $\delta>0$, loop induced annihilations into photons can be sufficiently large to distort CMB anisotropies at a measurable level, even though being too small to have to be included in the relic density calculation. We obtain for the associated annihilation cross section the expression
\begin{multline}
   (\sigma v_{\mathrm{lab}})_{\chi\bar{\chi} \to \gamma\gamma}
     = \sum_{l=e,\mu,\tau}  \frac{ 4 \alpha_{\mathrm{em}}^2  m_l^2 }{\pi (s - 2 m_\chi^2) } \frac{\alpha_\chi^P s+\alpha^S_\chi \left(s-4 m_\chi^2\right)}{(s-m^2_\phi)^2 + m_\phi^2 \Gamma_\phi^2} \\ \times \left\{\alpha^S_{ll} \left|1+(1 - \tau_l^{-2}) \asin^2\left(\tau_l\right) \right|^2 + \alpha_{ll}^P \left| \asin^2\left(\tau_l\right) \right|^2\right\},
     \label{eq:XSecPhot}
\end{multline}
with $\tau_l = \sqrt{s}/2 m_l$ and the fine-structure constant $\alpha_{\mathrm{em}}$. The forbidden region ruled out in this way based on the nBE calculation is shown in orange and in blue for the fBE treatment. Only the boundary of the cBE limit is marked in violet, as it is almost everywhere identical to the fBE result. These energy injection limits have been determined in Refs.~\cite{DAgnolo:2020mpt,Liu:2023kat} based on DM masses obtained with the nBE approach. Recalculating them based on the cBE and fBE approaches shows that the limits from annihilations into photons are actually more stringent and exclude a larger region of the parameter space. Included in red are also fBE projections from the CMB-S4 experiment~\cite{CMB-S4:2016ple}, which is expected to improve the limit on DM annihilation by a factor of two (dotted) to three (dashed).

For the muon channel of Fig.~\ref{fig:BPmuScan}, existing constraints from the electron beam-dump experiment E137 \cite{Bjorken:1988as,Marsicano:2018vin} on the light dark scalar are displayed in gray. In addition, projected limits from the experiments BDX~\cite{Bondi:2017gul,Marsicano:2018vin}, $\mathrm{M}^3$~\cite{Kahn:2018cqs} and NA64-$\mu$~\cite{Gninenko:2014pea,Chen:2018vkr} are shown. Here, we include, for the first time, the highest sensitivity limit from NA64-$\mu$ with $10^{13}$ muons on target (MOT) which is the goal the $\mathrm{M}^3$-experiment plans on achieving in phase two after starting out with comparably less $10^{10}$ MOT in phase one. The exclusion limits from beam-dump experiments for the benchmark with a non-vanishing pseudoscalar coupling are recasted from the associated limit with $\alpha^P_{\mu\mu}=0$ through the replacement $\alpha^S_{\mu\mu} \to \alpha^S_{\mu\mu}/2$ for every $m_\phi$. This is motivated by the fact that according to the improved Weizsacker-Williams approximation \cite{PhysRevD.8.3109}
the dominant contribution of the radiative production cross section $N\mu \to N\mu \phi$ of the scalar in presence of a nucleon $N$ in the target material is proportional to $\alpha^S_{\mu\mu}$.

For the tau channel of Fig.~\ref{fig:BPtauScan}, one existing constraint comes from the LEP measurement of the partial $Z$ decay width into tau leptons $\Gamma_{Z\to \tau^+\tau^-}=\SI{84.08\pm0.22}{\mega\electronvolt}$ \cite{ParticleDataGroup:2022pth} since it is sensitive to the process $Z\to \tau^+\tau^- \phi$ followed by a subsequent invisible decay of the new scalar into DM. To obtain a limit at the $2\sigma$ confidence level from this measurement, the new contribution to the $Z$ decay width is required to be less than two times the uncertainty of the measured $Z \to \tau^+\tau^-$ width, i.e. the upper limit $\Gamma_{Z\to \tau^+\tau^- \phi}\mathrm{BR}(\phi\to \chi\bar{\chi})  < \SI{0.44}{\mega\electronvolt}$ is applied.  For the purpose of constraining annihilations into tau leptons, we make here and in the following the simplifying assumption that $m_\chi\approx m_\tau$ within the computation of the branching ratio $\mathrm{BR}(\phi\to \chi\bar{\chi})$. $Z$ decays can also be probed with a better sensitivity at future electron-positron colliders like FCC-ee \cite{dEnterria:2016sca} or CEPC \cite{CEPCPhysicsStudyGroup:2022uwl} since these come with Giga-$Z$ (Tera-$Z$) options which means adjusting the beam energy to the $Z$-pole and producing $10^9$ ($10^{12}$) $Z$ bosons. Assuming similar efficiencies and acceptances of the future experiments for tau leptons as for electrons and muons, these colliders can probe the exotic $Z$ decay branching ratio $\mathrm{BR}(Z \to \tau^- \tau^+ \cancel{E})$ down to approximately $10^{-8}$ ($10^{-9.5}$) for a Giga-$Z$ factory (Tera-$Z$ factory) \cite{Liu:2017zdh}.
Another relevant experimental constraint comes from mono-photon searches at BaBar \cite{BaBar:2017tiz}, i.e. searches for a highly energetic monochromatic photon in association with missing energy. There are also projected limits for the same kind of search at BaBar's successor experiment Belle II \cite{Belle-II:2010dht} for integrated luminosities of $\SI{20}{\per\femto\barn}$ and $\SI{50}{\per\atto\barn}$. Both present and future constraints are recasted from mono-photon bounds on axion-like particles (ALPs) \cite{Dolan:2017osp}. To do so, we 
first consider the production cross section of an ALP $a$ in association with a photon given by
\begin{equation}
    \sigma_{e^+e^-\to a \gamma} = \frac{\alpha_{\mathrm{em}} g^2_{a\gamma\gamma}}{24 s^{7/2}} (s-m_a^2)^3 (s + 2 m_e^2) (s - 4 m_e^2)^{-\frac{1}{2}},
\end{equation}
where $g_{a\gamma\gamma}$ is the ALP-photon coupling and $m_a$ is the ALP mass. The same production cross section in our model, i.e. for the mediator $\phi$ instead of an ALP, is
\begin{equation}
     \sigma_{e^+e^-\to \phi \gamma} = \sum_l  \frac{2 \alpha_{\mathrm{em}}^3  m_l^2 }{3\pi  s^{7/2}} \frac{s + 2 m_e^2 }{s-m_\phi^2} (s-4 m_e^2)^{-\frac{1}{2}}\Big(  \alpha_{ll}^S |F_l^S(s)|^2 + \alpha_{ll}^P |F_l^P(s)|^2\Big),
\end{equation}
where the functional form of the form factors $F_l^S(q^2)$ and $F_l^P(q^2)$ due to the lepton loop are defined in App.~\ref{app:XSec}. We then recast the limit by solving the equation $\sigma_{e^+e^-\to a \gamma} =  \sigma_{e^+e^-\to \phi \gamma} \mathrm{BR}(\phi \to \chi\bar{\chi})$ for every $m_\phi$ at the collision energy $\sqrt{s} = \SI{10.58}{\giga\electronvolt}$ corresponding to the $\Upsilon(4S)$ resonance. Future $Z$-factories are able to perform the same mono-photon searches and can therefore put constraints on the branching ratio $\mathrm{BR}(Z\to \cancel{E}\gamma )$ \cite{Liu:2017zdh} which in our model corresponds to an upper limit on the product $\mathrm{BR}(Z\to\phi\gamma)\mathrm{BR}(\phi\to\chi\bar{\chi})$. The associated decay width is given by
\begin{equation}
     \Gamma_{Z \to \phi \gamma} = \sum_l \frac{3 \alpha_{\mathrm{em}} G_F (g^V_{Z,l})^2 m_l^2}{ \sqrt{2} \pi^3 m_Z ( m_Z^2 - m_\phi^2) } \Big(  \alpha_{ll}^S |F_l^S(m_Z^2)|^2 + \alpha_{ll}^P |F_l^P(m_Z^2)|^2\Big),
\end{equation}
with the Fermi constant $G_F$ and the vector part $g^V_{Z,l} =  -\frac{1}{4}+ \sin^2\theta_W$ of the $Z$-lepton coupling. \\

The implications of our new calculation on the available parameter space can be seen by comparing our results to the equilibrium results in Ref.~\cite{DAgnolo:2020mpt}. A very interesting outcome is that forbidden annihilations into muons for the case of a vanishing pseudoscalar coupling can now be entirely probed with the future CMB-S4 experiment alone\footnote{Note that Fig.~\ref{fig:BPmuScan} differs from the corresponding plot in Fig. 6 (left) of Ref.~\cite{Liu:2023kat} due to a potential error in their calculation of the mediator's width as a similar result is recovered by keeping $\Gamma_\Phi$ constant for the whole scan with a value corresponding to one far away from the resonance. In addition, the calculation of the limits on DM annihilation into photons provided in Figs. 6 and 7 of Ref.~\cite{Liu:2023kat} seem to be still based on the masses obtained with the number density treatment.}. Furthermore, there is a significant reduction in the experimentally viable parameter space of the model, with more regions available for the $\tau^+\tau^-$ final state. Many future experiments will be able to probe the remaining parts of the parameter space, such as NA64-$\mu$ and M$^3$ for the di-muon and the Giga-$Z$ and Tera-$Z$ experiments for the di-tau final state. The inclusion of the latter sensitivity limits is another novel aspect of this work and shows that almost the entire model parameter space can be probed in the near future. We note in passing that the discussed limits are more stringent for the $\alpha^P_{ll}=0$ case than they are for the $\alpha^P_{ll}=\alpha^S_{ll}$ case.

\section{Conclusion\label{sec:conclusion}}

We have studied the early kinetic decoupling effect for forbidden DM annihilations into SM leptons by means of the momentum-dependent Boltzmann equation. We carefully compared the resulting DM relic density with predictions obtained using the fluid approximation and the traditional number density approach for both, the full elastic collision term as well as the corresponding small-momentum transfer approximation, resulting in general in a significant increase of the DM relic abundance by more than an order of magnitude. From a technical side, we put particular emphasis on the analytical integration of all angular integrals appearing in the full elastic scattering collision term, as this possibility has not been explicitly addressed before in the context of DM. Along this line, we also highlighted improvements in the numerical strategy. With that, we derived from the relic density new experimental exclusion limits for the investigated model which are especially strong for the muon channel, however, by using the Fokker-Planck approximation instead of the full operator, since we found both to be in very good agreement in the relevant regions of the model parameter space. These results highlight again the necessity to take the early kinetic decoupling effect seriously and display the need to develop fast, reliable and general methods for the evaluation of full collision integrals in order to make the investigation of this effect in more complicated models like the Minimal Supersymmetric Standard Model feasible.

\appendix

\section{Cross sections}
\label{app:XSec}

\subsection{DM self-interaction}
As only the case $\alpha^S_\chi=0$ is covered in this work, the self-scattering cross sections are for brevity only displayed with the scalar coupling $\alpha^S_\chi$ set to zero. The cross sections are given by
\begin{multline}
   \sigma_{\chi\chi\to \chi\chi} = \frac{(\alpha_\chi^P)^2 \pi}{2 s} (3 s - 12 m^2_\chi + 5 m_\phi^2) \left\{\frac{1}{s - 4 m_\chi^2 + m_\phi^2} \right. \\ \left. +  \frac{2 m_\phi^2}{(s - 4m_\chi^2)(s - 4m_\chi^2 + 2 m_\phi^2)}\ln\left(\frac{m_\phi^2}{s - 4 m_\chi^2 + m_\phi^2}\right)\right\},
\end{multline}
\begin{multline}
   \sigma_{\chi\bar{\chi}\to \chi\bar{\chi}} = \frac{(\alpha_\chi^P)^2 \pi}{s} |D_\phi(s)|^2 \left\{\frac{m_\phi^2}{s - 4 m_\chi^2}  \left[2 |D_\phi(s)|^{-2} + s (m_\phi^2 - s)\right]\ln\left(\frac{m_\phi^2}{s - 4 m_\chi^2 + m_\phi^2}\right) \right. \\ \left. + \frac{m^2_\phi}{s - 4 m_\chi^2 + m_\phi^2} \left[s \left(\Gamma_\phi^2+4 m_\chi^2-2 m_\phi^2\right)+2 \left(\Gamma_\phi^2+m_\phi^2\right) \left(m_\phi^2-2
   m_\chi^2\right)\right] + s^2\right\},
\end{multline}
with the propagator $|D_\phi(s)|^2=1/((s - m_\phi^2)^2 + m_\phi^2 \Gamma_\phi^2)$.

\subsection{Loop-induced processes}
The form factors used to describe the processes $e^+ e^-\to \phi \gamma$ and $Z\to \phi \gamma$ are defined by
\begin{align}
    F_l^S(q^2) &= (q^2 - m^2_\phi)\left[2 + \left(q^2 +4
   m_l^2-m_\phi^2\right) C_0 \right]  +2 q^2 \left[ \Lambda \left(q^2,m_l,m_l\right)  - \Lambda
   \left(m_\phi^2,m_l,m_l\right)\right], \\ 
    F_l^P(q^2) &= (q^2 - m^2_\phi)^2 C_0\,,
\end{align}
with the Passarino-Veltman integral 
\begin{equation}
    C_0(0, m_\phi^2, q^2; m_l, m_l, m_l) = \frac{2}{q^2 - m_\phi^2} \left[\asin^2\left(\frac{m_\phi}{2 m_l}\right) - \asin^2\left(\frac{\sqrt{q^2}}{2 m_l}\right)\right],
\end{equation}
and the branch cut function
\begin{equation}
    \Lambda(p^2; m, m) = \sqrt{1- \frac{4 m^2}{p^2}}  \ln \left(\frac{\sqrt{p^2 \left(p^2-4 m^2\right)}+2 m^2-p^2}{2 m^2}\right).
\end{equation}

\section{The elastic scattering kernel}\label{app:D}

For the parametrization of the collision term 
\begin{align}
        C_{\mathrm{coll}}[f_a] = \frac{1}{16 (2\pi)^5 E_a g_1} \int \dd[3]{p_b} \dd[3]{p_1}  \dd[3]{p_2} \delta^{(4)}(p_a+p_b-p_1-p_2) \frac{|\mathcal{M}_{ab\to 1 2}|^2}{E_b E_1 E_2}   \mathcal{P}(f_a,f_b,f_1,f_2),
\end{align}
for a matrix element $|\mathcal{M}_{ab\to 1 2}|^2$ which depends only on one Mandelstam variable, we follow the strategy outlined in Refs. \cite{Hahn-Woernle:2009jyb,Hannestad:2015tea}. Without loss of generality, we choose this variable to be $t=(p_1 - p_a)^2=(p_b - p_2)^2$, define the corresponding three-momentum variable
\begin{equation}
    \vec{k} = \vec{p}_1 - \vec{p}_a=  \vec{p}_b - \vec{p}_2,
\end{equation}
and use it as reference direction to define the explicit coordinate system:
\begin{align}
    \vec{k} &= k  (0,0,1), \\
    \vec{p}_a &=   |\vec{p}_a| (0,\sin\eta,\cos\eta),  \\
    \vec{p}_b &= |\vec{p}_b| (\cos\varphi\sin\vartheta,\sin\varphi\sin\vartheta,\cos\vartheta). 
\end{align}
To avoid ambiguities, we differentiate in this section between a four-momentum $p$, its spatial component $\vec{p}$ and the associated absolute value $|\vec{p}|$.
The four-momentum $p_2$ is integrated out using four-momentum conservation
\begin{multline}
    \int \frac{\dd[3]{p_2}}{2 E_2} \delta^{(4)}(p_a+p_b-p_1-p_2)  = \Theta(E_a+E_b-E_1-m_2) \\ \times  \frac{1}{2 k |\vec{p}_b|} \delta\left(\cos\vartheta - \frac{|\vec{p}_b|^2 + k^2 + m_2^2 - (E_a+E_b-E_1)^2}{2 |\vec{p}_b| k}\right),
\end{multline}
where the remaining $\delta$-function sets $p_2$ on-shell. The absolute value of $\vec{p}_2$ is then fixed by energy conservation. 
The spatial components of $p_1$ are turned into the integration variable $\vec{k}$, giving
\begin{align}
    \int \frac{\dd[3]{p_1}}{2 E_1}  &= \int \dd[3]{k} \dd{E_1} \delta(E_1^2 - |\vec{k} + \vec{p}_a|^2 - m_1^2) \Theta(E_1 - m_1) \\
    &= \int \dd[3]{k} \dd{E_1} \frac{1}{2 k |\vec{p}_a|}  \delta\left(\cos\eta - \frac{E_1^2 - k^2 - |\vec{p}_a|^2 - m_1^2 }{2 |\vec{p}_a| k}\right) \Theta(E_1 - m_1).
\end{align}
After averaging over the direction of the incoming particle $(\int \dd{\!\cos\eta})/2$ and performing the trivial angular integrals, we find 
\begin{multline}
     C_{\mathrm{coll}}[f_a] = \frac{1}{128 \pi^3 E_a  |\vec{p}_a| g_1} \int \dd{E_1} \dd{E_b}  \dd{k}  \dd{\!\cos\vartheta} \dd{\!\cos\eta}   \delta\left(\cos\eta - \dots \right) \delta\left(\cos\vartheta - \dots \right)  \\ \times |\mathcal{M}_{ab\to 1 2}|^2   \mathcal{P}(f_a,f_b,f_1,f_2) \Theta(E_1 - m_1) \Theta(E_a+E_b-E_1-m_2) .
\end{multline}
The delta functions can be used to constrain the integration region of $k$ resulting in the limits
\begin{equation}
    k_-  \equiv \max(||\vec{p}_a|-|\vec{p}_1||,||\vec{p}_b|-|\vec{p}_2||)\leq k \leq  \min(|\vec{p}_a|+|\vec{p}_1|,|\vec{p}_b|+|\vec{p}_2|) \equiv k_+. 
\end{equation}
With that, we define the collision kernel 
\begin{equation}
    \Pi(E_a,E_b,E_1) = \Theta(k_+ - k_-) \int^{k_+}_{k_-} \dd{k} |\mathcal{M}_{ab\to 1 2}|^2,
\end{equation}
so that the final form of the full collision term reads 
\begin{equation}
     C_{\mathrm{coll}}[f_a] = \frac{1}{128 \pi^3 E_a  |\vec{p}_a| } \int^\infty_{m_1} \dd{E_1} \int^\infty_{\max(m_b,E_1-E_a+m_2)} \dd{E_b}  \Pi(E_a,E_b,E_1) \mathcal{P}(f_a,f_b,f_1,f_2).
\end{equation}
For a general $2$-to-$2$ matrix element that can be brought into the form 
\begin{equation}
    |\mathcal{M}_{ab\to 1 2}|^2 = c_0 + \frac{c_1}{\Delta_1 - k^2} + \frac{c_2}{(\Delta_2 - k^2)^2}\,,
\end{equation}
with $k$-independent coefficients $c_0$, $c_1$ and $c_2$, the kernel reads 
\begin{equation}
   \Pi(E_a,E_b,E_1) = \left[c_0 (k_+ - k_-) + c_1 I_1(\Delta_1,k_-,k_+)  + c_2 I_2(\Delta_2,k_-,k_+)\right]\Theta(k_+ - k_-),
\end{equation}
where we have introduced the two integrals
\begin{align}
    I_1(\Delta,a,b) = \int_a^b \dd{x} \frac{1}{\Delta - x^2} =  %
    \begin{cases}
     \frac{1}{2\sqrt{\Delta}} \left[\ln \left(\frac{\sqrt{\Delta}-a}{\sqrt{\Delta}-b}\right)+\ln
   \left(\frac{b+\sqrt{\Delta}}{a+\sqrt{\Delta}}\right)\right]&, \Delta>0 \\
   \frac{1}{b} - \frac{1}{a}&, \Delta =0 \\
     \frac{1}{\sqrt{-\Delta}}\left[\mathrm{atan}\left(a/\sqrt{-\Delta}\right) -\mathrm{atan}\left(b/\sqrt{-\Delta}\right)\right]&, \Delta <0 
    \end{cases}
\end{align}
as well as
\begin{align}
    I_2(\Delta,a,b) = \int_a^b \dd{x} \frac{1}{(\Delta - x^2)^2} =  %
    \frac{1}{2\Delta} \left(\frac{a}{a^2-\Delta} - \frac{b}{b^2-\Delta} + I_1(\Delta,a,b) \right).
\end{align}
For the elastic scattering matrix element in our model
\begin{equation}
   |\mathcal{M}_{\chi l \to \chi l}|^2 =  \frac{64\pi^2}{(t-m_\phi^2)^2} \Big(\alpha^P_{\chi} t+ \alpha^S_\chi \left(t-4 m_\chi^2\right)\Big) 
   \Big(\alpha^P_{ll} t+ \alpha^S_{ll} \left(t-4  m_l^2\right)\Big),
   \label{eq:M2scatter}
\end{equation}
one has $\Delta = (E_a - E_1)^2 - m_\phi^2$ 
and the coefficients $c_0$, $c_1$ and $c_2$ are given by
\begin{align}
c_0&=64\pi^2(\alpha_\chi^{S}+\alpha_\chi^{P})(\alpha_{ll}^{S}+\alpha_{ll}^{P}), \\
c_1&=128\pi^2\Big\{\Big(\alpha_{ll}^S+\alpha_{ll}^P\Big)\Big(\alpha_\chi^S(m_\phi^2-2m_\chi^2)+\alpha_\chi^P m^2_\phi\Big)-2\alpha_{ll}^S(\alpha_{\chi}^S+\alpha_{\chi}^P)m_{l}^2\Big\}, \\
c_2&=64\pi^2\Big(\alpha_\chi^S(m_\phi^2-4m_\chi^2)+\alpha_\chi^P m^2_\phi\Big)\Big(\alpha_{ll}^S(m_\phi^2-4m_{l}^2)+\alpha_{ll}^P m^2_\phi\Big).
\end{align}

\acknowledgments
We would like to thank Di Liu for correspondence regarding limits from future $Z$-factories. The research of AA, MK and LPW was supported by the DFG through the Research Training Group 2149 ``Strong and Weak Interactions - from
Hadrons to Dark matter''. The work by AA and MK was also funded by the BMBF under contract 05P21PMCAA and grant KL 1266/10-1. 
Matrix elements and cross sections have been computed with the help of \texttt{FeynCalc} \cite{Shtabovenko:2016sxi,Shtabovenko:2020gxv} and \texttt{Package-X} \cite{Patel:2015tea,Patel:2016fam}.

\bibliography{references.bib}

\end{document}